\pdfoutput=1
\documentclass[aps,prl,twocolumn,showpacs,amssymb,superscriptaddress,floatfix]{revtex4-1} 

\usepackage{amsmath}

\usepackage{graphicx}
\usepackage{float}

\usepackage{relsize}

\usepackage{color}
\usepackage{verbatim}
\usepackage{natbib}

\usepackage{afterpage}

\newcommand\blankpage{
    \null
    \thispagestyle{empty}
    \addtocounter{page}{-1}
    \newpage
    }

\newcommand{\be}{\begin{equation}}
\newcommand{\ee}{\end{equation}}
\newcommand{\cu}{\gamma_{SB}}
\newcommand{\tcu}{\tilde{\gamma}_{SB}}
\newcommand{\gu}{\gamma_{BB}}

\begin{document}

\title{Swift heat transfer by fast-forward driving in open quantum systems}

\author{Tamiro Villazon}
\affiliation{Department of Physics, Boston University, 590 Commonwealth Ave., Boston, MA 02215, USA}
\author{Anatoli Polkovnikov}
\affiliation{Department of Physics, Boston University, 590 Commonwealth Ave., Boston, MA 02215, USA}
\author{Anushya Chandran}
\affiliation{Department of Physics, Boston University, 590 Commonwealth Ave., Boston, MA 02215, USA}	
\begin{abstract}
\noindent Typically, time-dependent thermodynamic protocols need to run asymptotically slowly in order to avoid dissipative losses. By adapting ideas from counter-diabatic driving and Floquet engineering to open systems, we develop fast-forward protocols for swiftly thermalizing a system oscillator locally coupled to an optical phonon bath. These protocols control the system frequency and the system-bath coupling to induce a resonant state exchange between the system and the bath.
We apply the fast-forward protocols to realize a fast approximate Otto engine operating at high power near the Carnot Efficiency. Our results suggest design principles for swift cooling protocols in coupled many-body systems.
\end{abstract}
	
	\maketitle

\section{Introduction}

\noindent
Fast and efficient heat transfer using small quantum systems plays an important role in microscopic heat engines~\cite{Vinjanampathy,Kosloff1,Levy,Harbola,Linden}, reservoir engineering~\cite{Koch}, and many-body state preparation~\cite{Chandra,Bohn,Verstraete}.
There are now many experimental platforms, such as NV centers in diamond~\cite{Schirhagl,Klatzow}, trapped ions~\cite{Rossnagel,Maslennikov, Blatt}, and superconducting circuits~\cite{Wendin, Pekola,Fornieri}, capable of preparing and coherently manipulating small quantum systems.
An important experimentally relevant question, which we address in this article, is how to achieve swift and efficient heat transfer with limited control of system and system-bath parameters.

There is generally a trade-off between control speed and efficiency~\cite{Reif,Kolodrubetz}. Reversible processes attain maximal efficiency; however these need to run asymptotically slowly to remain in instantaneous equilibrium. Slow driving can be impractical or even prohibitive in real applications which need to run in finite time to avoid decoherence or generate power. On the other hand, fast driving typically forces the system out of instantaneous equilibrium and results in dissipative losses that reduce efficiency. 

In isolated systems, fast reversible processes can be realized using shortcuts to adiabaticity, an umbrella term used for counter-diabatic (CD) and fast-forward (FF) protocols. CD protocols suppress transitions between the instantaneous eigenstates of a target driven Hamiltonian $H(t)$ by evolving the system with a modified Hamiltonian $H_{CD}(t)$ ~\cite{Demirplak1,Demirplak2,Demirplak3,Berry,delCampo2,Muga,Kolodrubetz}. A similar strategy for suppressing transitions is implemented in closely related superadiabatic protocols~\cite{Zhou}. Usually, the CD Hamiltonians require access to non-local controls not present in the original Hamiltonian. FF protocols, on the other hand, only modulate the couplings present in the original Hamiltonian to attain the desired adiabatic final state~\cite{delCampo2,Muga,Kolodrubetz, Masuda,Torrontegui,Bukov1}. These FF protocols are related to CD protocols by time-dependent unitary transformations~\cite{Bukov1}. Several works have used such CD and FF protocols to speed up the adiabatic parts of various thermodynamic cycles~\cite{Tu, Deng, delCampo1, Beau, Kosloff2}.

\begin{figure}[b]
\centering
\includegraphics[width=1.06\columnwidth]{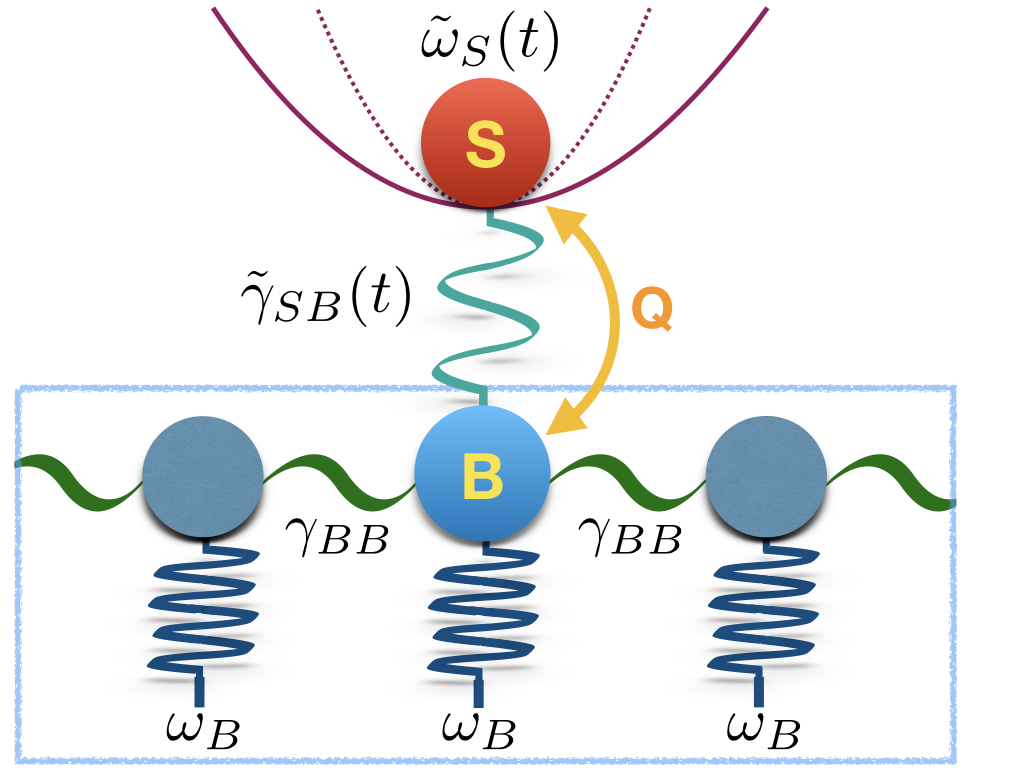}
\caption{\textbf{Schematic depiction of a small quantum system coupled to an optical phonon bath.} 
A system oscillator S with tunable frequency $\tilde{\omega}_S(t)$ is coupled to a one-dimensional optical phonon bath with central frequency $\omega_B$ via the bath oscillator B. 
We derive FF protocols that resonantly exchange the states of the S and B oscillators by suitably modulating the system frequency $\tilde{\omega}_S(t)$ and the system-bath coupling $\tcu(t)$. 
These protocols swiftly thermalize S through a rapid heat exchange Q with the bath and far out-perform unassisted protocols.}
\label{fig:model}
\end{figure}

In this article, we extend FF driving to a small open system. We present new FF protocols which realize an efficient energy exchange between the system and its environment and swiftly thermalize the system. These protocols are constructed using a tractable model for an oscillator system locally coupled to a non-Markovian optical phonon bath (Fig.~\ref{fig:model}). We use these protocols to design a fast (high-power) heat engine operating near the Carnot efficiency. Importantly, these protocols can be experimentally realized, as they only demand control over system parameters and the system-bath coupling.  

The ideas of shortcuts to adiabaticity were recently generalized to speed up equilibration and isothermal processes in open systems~\cite{Martinez, Chupeau, Dann, Li,Patra,Boyd, Vacanti}.
Such protocols assume Markovian baths and are effective when the protocol duration is much longer than the bath relaxation time. 
They however often lead to dissipative losses which increase with the driving speed. 
Our results are complementary in three respects.
First, the bath in our setup has a narrow bandwidth and is not Markovian. To capture the non-Markovian effects, we model the system+bath microscopically as a Hamiltonian system~\cite{Weiss}. 
Second, our FF protocols are most effective when the protocol duration is much shorter than the relaxation time of the bath.
Strikingly, the performance of these FF protocols is a non-monotonic function of the protocol duration, suggesting that the Markovian protocols and our FF protocols are not limiting behaviors of a single general protocol. 
Finally, unlike the Markovian protocols, the heat dissipated during our FF protocols remains bounded at all driving speeds.

\section{Model}
\noindent
We model the small quantum system as a tunable harmonic oscillator S with Hamiltonian $H_{S}(t) = \frac{1}{2} P^2 + \frac{1}{2} \,\tilde\omega_{S}^2(t) X^2 $,
which is connected to an optical phonon bath via the bath oscillator B (see Fig.~\ref{fig:model}). 
The complete Hamiltonian for the system and bath is given by:
\be\label{H}
H = H_S(t) + H_{bath} + H_{SB}(t) 
\ee
where 
\be\label{Hbath}
H_{bath}=\sum_{j=1}^{N} \,\bigg[\frac{1}{2} p_{j}^2 + \frac{1}{2}  \omega_{B}^2 x^2_{j} - \,\gu \, \omega_B^2\, x_{j} \,x_{j+1} \bigg]
\ee
describes the bath of $N$ optical phonons with central frequency $\omega_B$. The B oscillator is indexed by $j=B$, and $H_{SB}(t) =  -\tcu(t) \,\omega_B^2 \,x_B\,X$ describes the tunable interaction between S and B. The bare S-B coupling strength when $H_{SB}$ is not varied in time is denoted by $\cu$. The bare coupling $\cu$ can be also viewed as a boundary condition for $\tcu(t)$ at the beginning and at the end of a protocol. We work in the regime $\gu, \cu \ll 1$, in which all oscillators interact weakly with one another. 

We study different driving protocols of the S oscillator frequency $\tilde{\omega}_S(t)$ and S-B coupling strength $\tcu(t)$. 
In unassisted (UA) driving, the system's frequency is varied in time as $\omega_S(t)$, while the S-B coupling is time independent $\tcu(t) = \cu$. 
Assisted fast-forward (FF) protocols modulate both couplings in time, targeting the same final state as in an adiabatic UA protocol.

The target ramps of $\omega_S(t)$ sweep across the bandwidth of the bath frequencies. 
We define the dimensionless detuning parameter
\be
\lambda(t)=\frac{\omega_S^2(t)-\omega_B^2}{ \omega_B^2},
\ee
so that S is resonant with the central frequency of the bath at $\lambda=0$. In a target ramp, $\lambda(t)$ is initialized with a value $\lambda_i$ at $t=t_i$ and driven to a final value $\lambda_f$ at $t=t_f$. For concreteness, we consider a linear ramp $\lambda(t)$ which rounds-off sufficiently smoothly at the ramp boundaries. The ramp duration is denoted by $\tau_p = t_f - t_i.$ All FF protocols in this work can be parameterized by $\lambda(t)$, enabling direct comparison with UA protocols.

Since $H$ is quadratic, our analysis is valid for both quantum and classical oscillator systems. For concreteness, we use the language of quantum mechanics. Thus, symbols such as $X$ or $p_B$ are to be understood as operators. The occupation number operator of oscillator/mode $a$ is denoted by $n_a$. Expectation values such as $\langle n_a (t) \rangle$ are with respect to the state at time $t$. We set $\hbar=1, k_B=1$.

\section{The two oscillator subsystem}

\begin{figure}[t]
\centering
 \includegraphics[width=1.05\columnwidth,trim=4 4 4 4,clip]{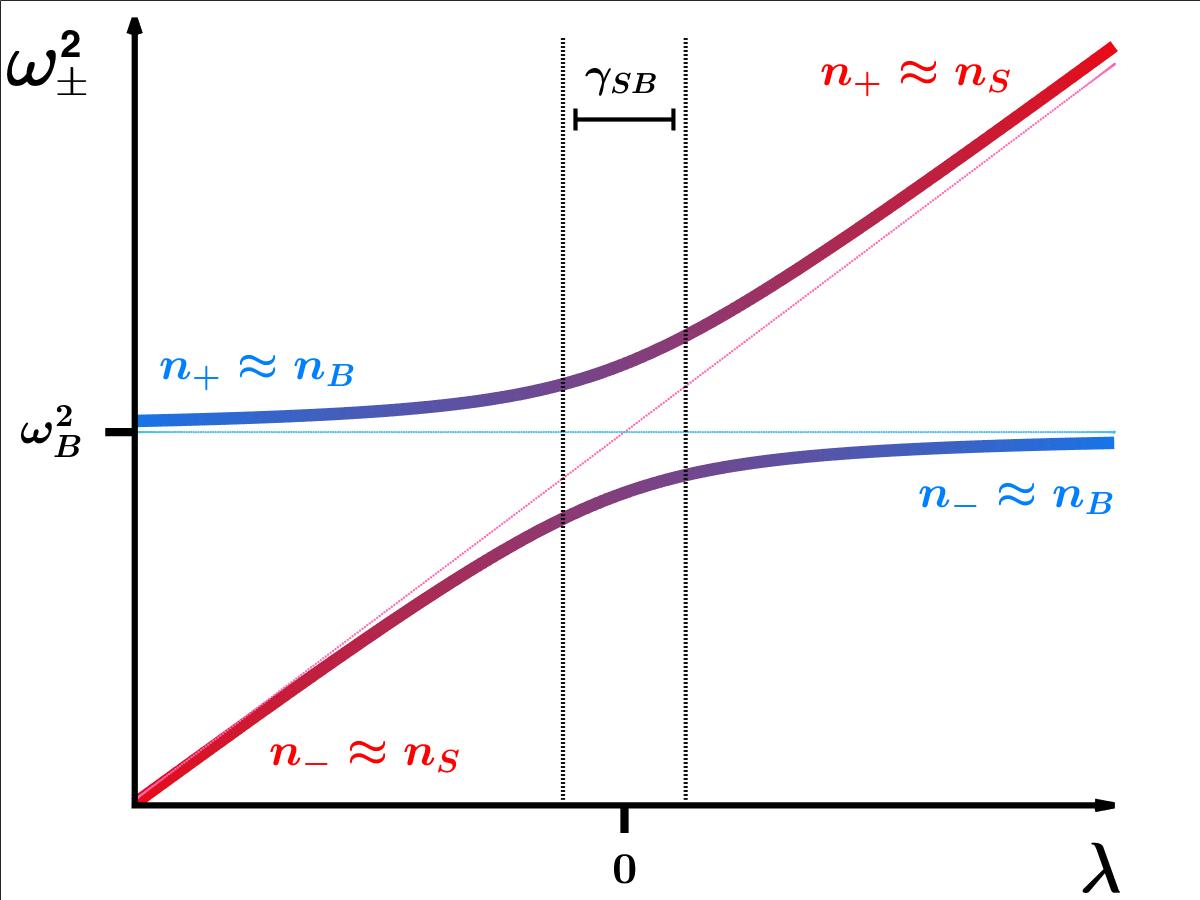}
\caption{\textbf{Adiabatic driving through resonance switches the instantaneous occupations of two coupled oscillators. } Schematic plot of the normal mode frequencies $\omega_{\pm}^2$ as a function of the detuning $\lambda$ for the two oscillator system. Red and blue denote the approximately unhybridized S and B oscillators respectively. 
Far from the resonance ($|\lambda| \gg \cu$), S and B are approximately distinct normal modes with occupation numbers ${n}_{S}$ and $n_B$. Near resonance ($|\lambda| \lesssim \cu$), S and B hybridize. An adiabatic or CD protocol suppresses transitions between the normal modes and induces an occupation switch \big($ {n}_S \leftrightarrow  {n}_B $\big) between S and B.
 }
\label{fig:exchange}
\end{figure}

\noindent
When $\gu=0$, the S and B oscillators decouple from the rest of the bath.
The dynamics is thus determined by the Hamiltonian $H_0 = H_S(t)+H_B+H_{SB}$ for two coupled harmonic oscillators, where  $H_B = \frac{1}{2}p_B^2 + \frac{1}{2}\,\omega_{B}^2\, x_B^2 $ .\\

\noindent
\textbf{The Unassisted (UA) Ramp.} 
Fig.~\ref{fig:exchange} depicts the frequencies $\omega_{\pm}$ of the instantaneous normal modes as a function of $\lambda$.
Far from resonance $|\lambda(t)/\cu| \gg 1$, the S and B oscillators weakly hybridize and the normal modes are either completely of S (red) or B (blue) character.
In the resonance region $|\lambda(t)/\cu| \lesssim 1$ on the other hand, the normal modes are approximately equal weight superpositions of the S and B modes.
As $\lambda(t)$ is tuned across resonance, the instantaneous normal mode of S character evolves continuously across the resonance region to the normal mode with B character, and vice-versa.

An adiabatic ramp induces no transitions between the instantaneous eigenstates of $H_0(t)$. 
As the normal modes preserve their occupation numbers ${n}_{\pm}$, the occupation numbers of S and B exchange \big(${n}_S \leftrightarrow {n}_B$\big) across resonance. 
In particular, if we prepare B in a thermal distribution at temperature $T$, then S will acquire this distribution when driven slowly enough through resonance. 
This exchange-induced thermalization is reversible; that is, the S+B system comes back to its initial state if the direction of the ramp is reversed. 

At finite ramp rates $\dot{\lambda}$, there are two classes of excitations between the instantaneous energy levels of $H_0(t)$. The first class consists of number-conserving exchanges of energy quanta between the normal modes of $H_0(t)$. These exchanges occur near resonance and are important when $\dot{\lambda}$ becomes comparable to the scale $\omega_B\,\cu^2$. This is analogous to a two-level Landau-Zener (LZ) problem where the onset of non-adiabatic transitions is marked by a speed scale proportional to the square of the interaction gap \cite{Shevchenko,AP1}. The second class consists of quanta pair creation/annihilation and becomes important when the ramp speed $\dot{\lambda}$ is comparable to the larger scale of $\omega_B$ [SI Text]. Both processes induce diabatic transitions in the instantaneous eigenbasis of $H_0(t)$ and reduce the fidelity of the S-B exchange.  \\

\noindent
\textbf{Counter-Diabatic (CD) Driving.} To prevent diabatic transitions at any detuning speed $\dot{\lambda}$, we engineer a CD Hamiltonian:

\be
H_{\rm CD}(t)=H_0(\lambda(t))+ {\dot \lambda} \,\mathcal A(t)
\label{H_CD}
\ee
where the gauge potential $\mathcal A(t)$ is found to be [SI Text]:
\begin{align} \nonumber
\label{Gauge_pot}
\mathcal A(t) \approx &- \frac{1}{ 4(1+\lambda(t))}\,(X\,P+P\,X)  \\ &+ \frac{\cu}{\lambda^2(t) + 4 \cu^2}\, (X \,p_B -x_B \,P) 
\end{align}

The first term in Eq.~\eqref{Gauge_pot} represents the gauge potential in the absence of the system-bath coupling (see e.g. Ref. \cite{Deffner}). It is responsible for suppressing diabatic transitions in the S oscillator. The second term dynamically exchanges the S and B oscillator states across resonance, thus preserving normal mode occupation numbers [SI Text]. Near resonance, this term scales as $\cu^{-1} \gg 1$. It enhances the interaction between S and B to speed up the exchange at finite ramp speeds. The terms neglected in Eq.~\eqref{Gauge_pot} are suppressed by higher powers of $\cu$ [SI Text] and do not qualitatively change the following discussion. 

The CD protocol given by Eq.~\eqref{H_CD} realizes transitionless driving for arbitrary $\lambda(t)$. However, it requires new couplings ($X\,P$, $P\,X$, $X\,p_B$, $x_B\,P$) which are not present in the original Hamiltonian $H_0$ and which are hard to realize experimentally. \\

\noindent
\textbf{Fast-Forward (FF) Driving.} FF Hamiltonians can generally be obtained by unitary rotations of CD Hamiltonians~\cite{Bukov1}: $H_{FF} = U^{\dagger} \,H_{CD} U - i\, U^{\dagger} \partial_t U$. Here $U$ is a unitary transformation that enforces that $H_{FF}$ has the same form as $H_0$, but with different time-dependence of the S frequency $\tilde{\omega}_S(t)$ and S-B coupling $\tcu(t)$. In addition, for FF protocols to robustly attain the same target state as CD, $U$ must coincide with the identity and have vanishing time derivatives at the protocol boundaries \cite{Ness,Kolodrubetz, Torrontegui}; [SI Text].

When $\dot{\lambda} \ll \omega_B$, we construct a simple FF protocol in a rotating wave (RW) approximation which ignores pair creation/annihilation processes (see Methods):
\be \label{HRW}
H_{FF}^{RW}=\frac{P^2}{2} + \frac{\tilde{\omega}_{S}^2(t)}{2} X^2 +  \frac{p_B^2}{2} + \frac{\omega_{B}^2}{2} x_B^2 -  \tcu(t)\omega_B^2\,x_B X,
\ee
where 
\begin{align} 
&\tilde{\omega}^2_S(t)=\omega_B^2(1+\lambda(t))+ 2\,\omega_B\ \frac{d}{dt}\tan^{-1}\bigg[\frac{\,\dot \lambda(t) \,\omega_B^{-1}}{ \lambda^2(t)+4\,\cu^2}\bigg],\label{RW2} \\ 
&\tcu(t) = \cu\, \sqrt{ 1 + \bigg[\frac{2\,\dot \lambda(t) \, \omega_B^{-1}}{  \lambda^2(t)+4\,\cu^2}\bigg]^2 }.\label{RW3}
\end{align}

In a real setup, $\tilde{\omega}_S^2(t)$ and $\tcu(t)$ are physical control knobs. Contrary to UA and CD protocols, $\lambda(t)$ is no longer the physical detuning, except at the protocol boundaries. Rather, $\lambda(t)$ should be understood as a free function parameterizing a family of FF protocols, with boundary conditions: $\lambda(t_i) = \lambda_i$, $\lambda(t_f) =\lambda_f$, and $\dot\lambda(t_{i,f})=0$. The latter condition ensures that FF achieves the target adiabatic state. We also impose $\ddot\lambda(t_{i,f})=0$ to ensure $H_{FF}^{RW} = H_0$ at the boundaries and stabilize the final state after the ramp. Given a target UA protocol $\lambda(t)$ satisfying these conditions, Eqs.~\eqref{RW2} and \eqref{RW3} show how it must be modulated to realize the RW-FF protocol. 

\begin{figure}[t]
\centering
\includegraphics[width=1.06\columnwidth]{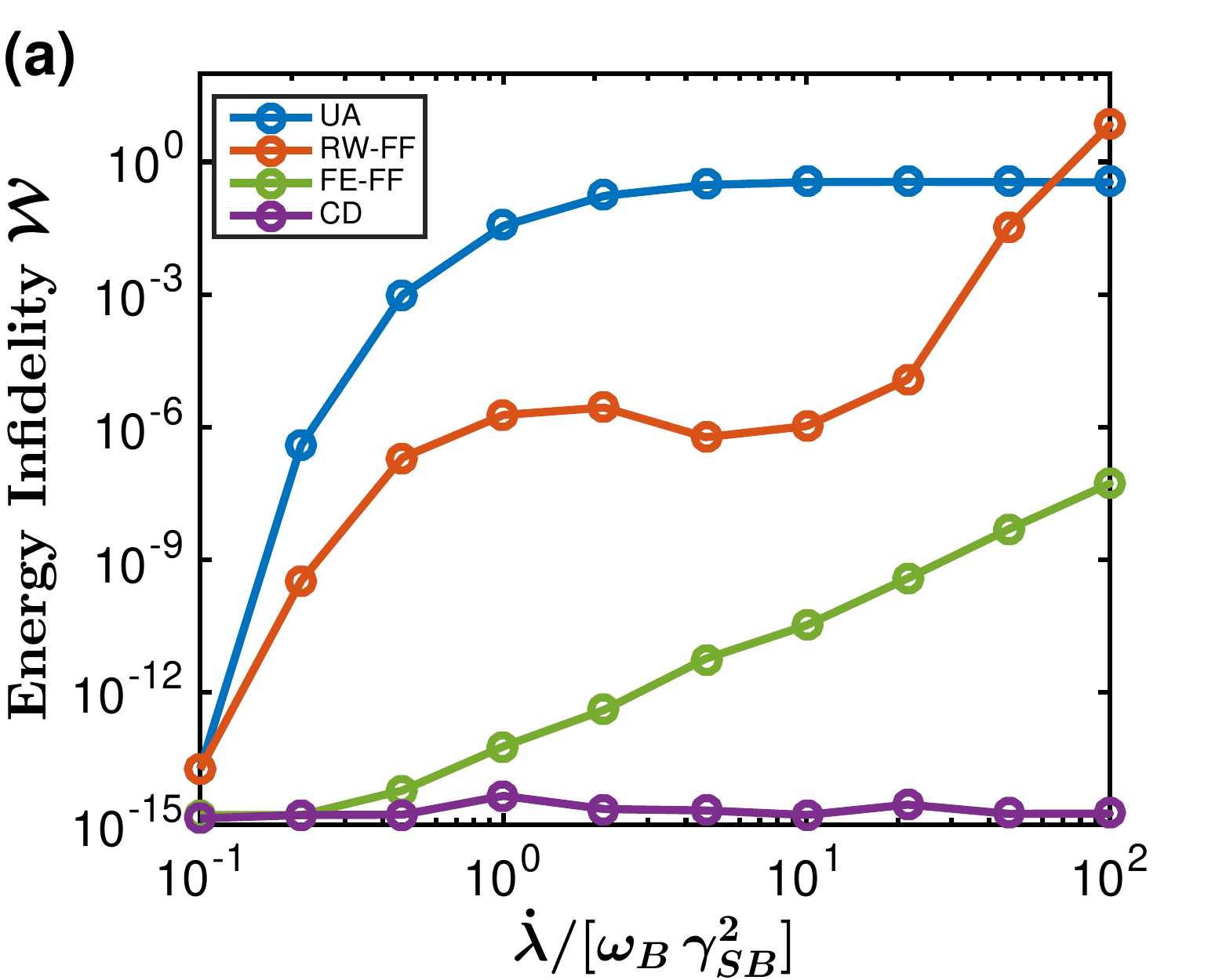}
\includegraphics[width=1.00\columnwidth]{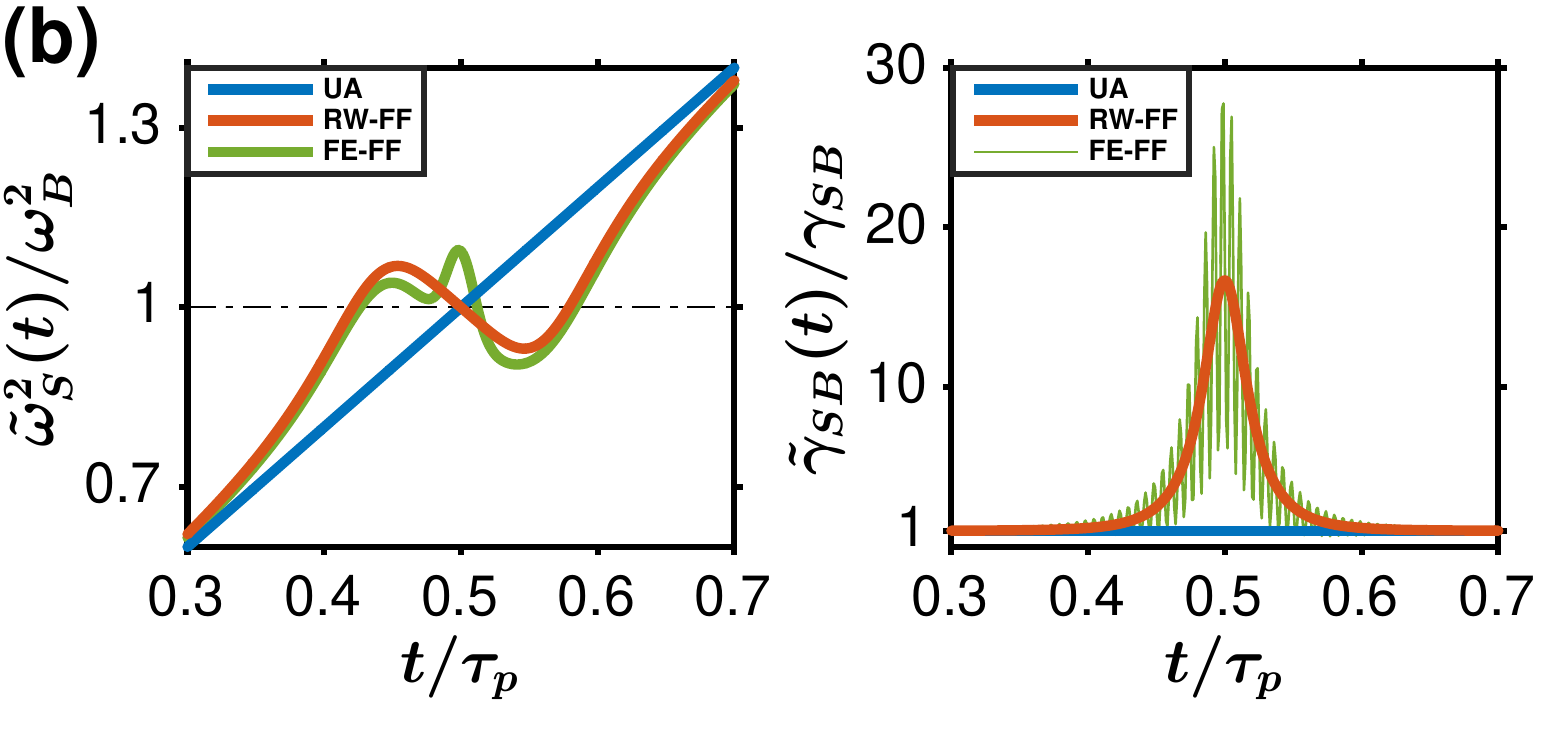}
\caption{\textbf{FF protocols suppress diabatic transitions by controlling $\tilde{\omega}_S(t)$ and $\tcu(t)$.} (a) The energy infidelity $\mathcal W$ (c.f. Eq.~\eqref{Efid}) vs normalized ramp speed $\dot{\lambda}/[\omega_B\, \cu^2]$ for several protocols. The rotating-wave
(RW-FF) protocol outperforms UA when $\dot{\lambda}\ll\omega_B$, while the Floquet-Engineered (FE-FF) protocol outperforms UA and RW-FF at all speeds. The exact CD drive reproduces an adiabatic protocol to within numerical accuracy. (b) Time-modulations of $\tilde\omega_S(t)$ (left) and $\tcu(t)$ (right) over the ramp period $\tau_p$ in FF driving. Simulation parameters: (a - b) $\lambda_i = -0.67$, $\lambda_f = 0.67$, $\omega_B = 3$, $\cu = 0.02$, $|n_{-}(\lambda_i),n_{+}(\lambda_i)\rangle = |3,1\rangle $; (a) $\Omega=480$; (b) $\Omega = 1$.}
\label{fig:fidelity}
\end{figure}

Fig.~\ref{fig:fidelity}b shows the time-modulations $(\ref{RW2})$ and $(\ref{RW3})$ for a ramp across resonance (only the linear part of the ramp near resonance is shown). To achieve the S-B state exchange, the RW-FF protocol non-monotonically modulates $\tilde{\omega}_S^2(t)$ to keep S resonant with B for a longer time span than UA, while simultaneously enhancing $\tcu(t)$ in the resonance region ($|\lambda(t)| \lesssim \cu$). 

One can improve upon the RW-FF approximation and design an exact local FF protocol which can be implemented to arbitrary precision by a high-frequency Floquet drive of $\tcu(t)$. The precision error is set by the period $2\pi/\Omega$ of the drive. In the resulting Floquet-Engineered Fast-Forward (FE-FF) Hamiltonian $H_{FF}^{FE}$, $\tilde{\omega}_S(t)$ and $\tcu(t)$ become complicated functions of time in comparison to their bare counterparts, as shown in Fig.~\ref{fig:fidelity}b. Similar to Eqs.~\eqref{RW2} and \eqref{RW3}, $\tilde\omega_S(t)$ is non-monotonic, while the new $\tcu(t)$ enhances the S-B interaction to effect the S-B state exchange. Now however, $\tcu(t)$ has an added high-frequency periodic modulation $\propto \Omega \,\cos(\Omega \,t)$ needed to indirectly control the bath frequency $\omega_B$ and suppress transitions at any speed $\dot\lambda$. In the Methods section, we outline the construction of this protocol, and give a thorough treatment in [SI Text]. 

  It was shown in Refs.~\cite{Boyers, Petiziol, Sun} that approximate FF protocols can be designed using high-frequency periodic driving in specific setups. Recently, a high-frequency FE-FF protocol was also realized in an experiment with NV centers in diamond to achieve high-fidelity state preparation in a qubit \cite{Boyers}. The advantages of this kind of approach range from experimental viability to robustness against environmental noise~\cite{Viola}. \\

\noindent
\textbf{Protocol Comparison:} We compare the performance of the UA, RW-FF, and FE-FF protocols by measuring the energy infidelity, $\mathcal{W}$. This quantity is a proxy for diabatic transitions, depending only on measurable quantities such as mean energy and energy variance:
\be\label{Efid}
\mathcal{W} = \frac{1}{\omega_B^2}\big((E-E_{ad})^2+\sigma_E^2\big).
\ee
Here, $E=\langle H_0 \rangle$ and $\sigma_E^2=\langle H_0^2\rangle-E^2$ are the mean total energy and energy variance of the S+B subsystem at the end of the protocol. $E_{ad}$ is the mean total energy in the final state for an adiabatic UA protocol. All protocols are initialized in an eigenstate of $H_0(\lambda_i)$.

 Fig.~\ref{fig:fidelity}a shows the energy infidelity $\mathcal{W}$ as a function of the normalized ramp speed $\dot{\lambda}/[\omega_B\,\cu^2]$ for various protocols. The exact CD protocol realizing a perfect adiabatic process is shown for reference. For this protocol, $\mathcal{W}= 0$ within numerical accuracy. In contrast, $\mathcal{W}$ dramatically rises for $\dot{\lambda}>\omega_B\,\cu^2$ in the UA protocol. The RW-FF protocol shows a substantial improvement over UA, suppressing $\mathcal{W}$ by several orders of magnitude in the regime $\omega_B\,\cu^2\lesssim\dot{\lambda}\ll \omega_B$. At sufficiently large speeds, RW-FF is not effective because the rotating wave approximation breaks down when $\omega_B$ becomes a relevant time scale. The FE-FF protocol outperforms the UA and RW-FF protocols at all speeds. We observe such improvement whenever $\Omega$ is the largest frequency scale. In this regime $\mathcal{W} \sim \Omega^{-1}$, so that FE-FF approaches a perfect adiabatic protocol as $\Omega\to\infty$ [SI Text].

\section{The Many-Oscillator Environment}

\noindent
When all the oscillators in the phonon bath are coupled ($\gu>0$), the normal modes of the bath have frequencies within the bandwidth $[\omega_B(1-\gu),\omega_B(1+\gu)]$ around the central frequency $\omega_B$. Then $\gu$ sets the internal relaxation rate of the bath. For the UA protocol, this scale competes with the timescale set by $\cu$ for the S-B interaction. When $\gu\ll \cu$, the dynamics are qualitatively similar to the $\gu=0$ case described above. When $\gu\gtrsim\cu$, S interacts with multiple bath normal modes when $\omega_S(t)$ lies within the bandwidth. These interactions thermalize S and give rise to a reversible isothermal process when $\omega_S(t)$ is slowly ramped across the bandwidth. Fast unassisted ramps, however, fail to thermalize S because they leave no time to exchange sufficient energy with the bath. 
 
\begin{figure}[htp]
    \centering
    \includegraphics[width=1.07\columnwidth]{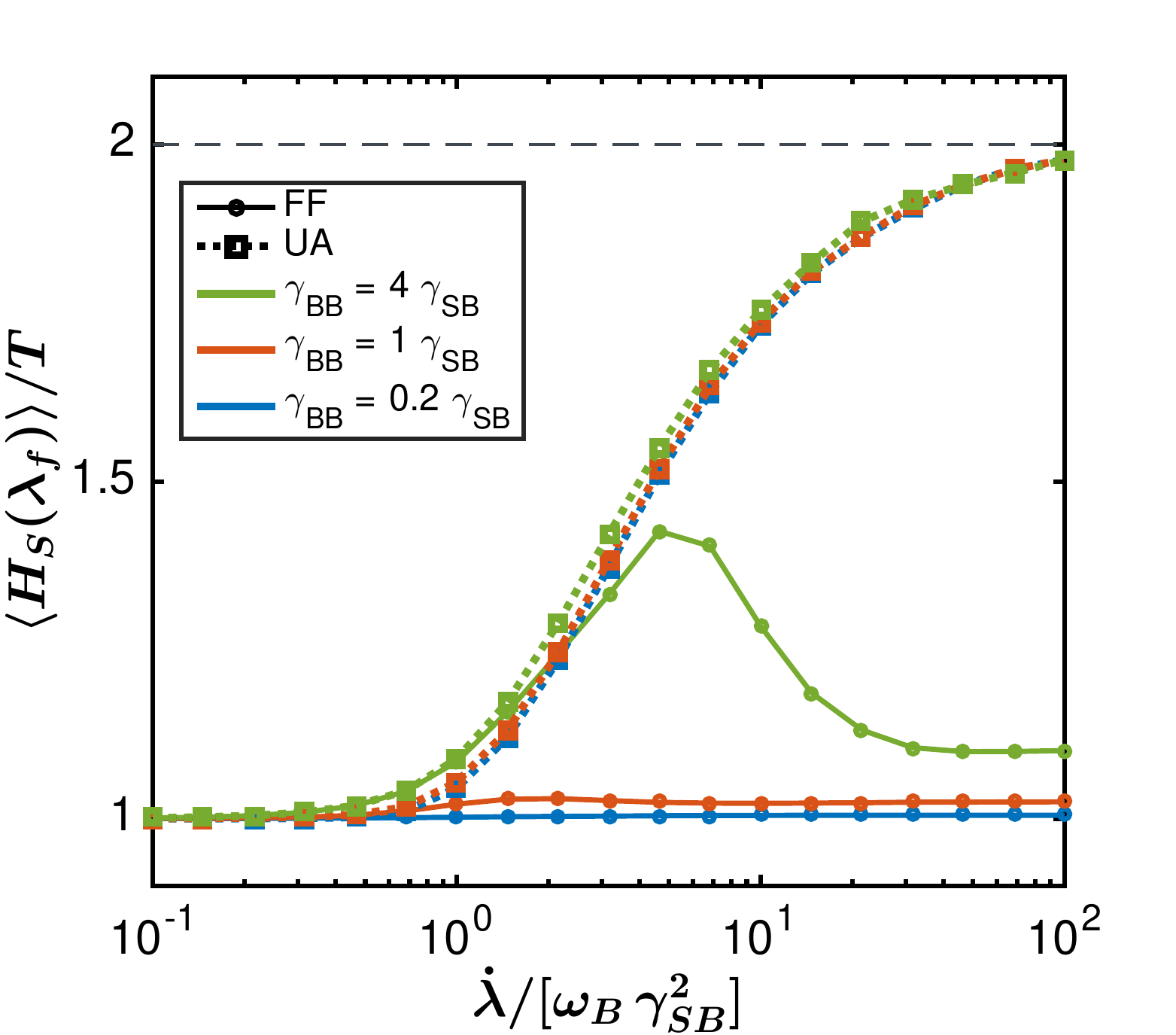}
    \caption{\textbf{FF driving induces swift thermalization with a phonon bath.} Plot of the system's normalized average energy $\langle H_S(\lambda_f)\rangle/T$ at the end of a ramp across the bath's bandwidth as a function of normalized ramp speed $\dot{\lambda}/[\omega_B\, \cu^2]$. Here $T$ is the temperature of the bath. S is prepared at $\lambda_i = - 2\,\gu$ in a ``hot'' state and ramped to $\lambda_f = 2 \gu$. The ``hot" state is chosen to yield a final temperature $2\,T$ (black dotted line) if S does not interact with the bath. Although the UA and FF protocols thermalize S at slow speeds, only the FF protocols thermalize S to temperatures near $T$ at fast speeds. Simulation parameters: $N=100$, $\omega_B=\sqrt{2}$, $\cu=0.025$.} 
    \label{fig:isothermality}
\end{figure}

The FF protocols developed in the last section thermalize S through a reversible S-B state exchange at $\lambda=0$. Fig.~\ref{fig:isothermality} shows the final temperature of S for ramps across the bandwidth of a bath at temperature $T$. 
S is initially prepared with mean occupation $2T/\omega_S(t_f)$, so that its final temperature is $2\,T$ for adiabatic ramps in the absence of the bath. As S effectively does not interact with the bath in fast UA ramps, its final temperature is $2T$ in Fig.~\ref{fig:isothermality}.  
In contrast, the FF protocols yield a final temperature near $T$ as $\dot{\lambda} \to \infty$.
When $\dot{\lambda} \ll \omega_B\,\cu^2$, S is in instantaneous equilibrium at all times in all the protocols. Consequently, all protocols result in a perfect isothermal process at temperature $T$. 

In Fig.~\ref{fig:isothermality}, the final temperature of S monotonically increases for $T$ to $2T$ as a function of $\dot{\lambda}$ in UA protocols. The FF curves, on the other hand, are non-monotonic; specifically, they approximately follow the UA curve up to some value of $\dot{\lambda}$ and then peel off towards $\langle H_S(\lambda_f) \rangle \approx T$.
We expect that the FF protocols become effective for thermalization only when the duration of the resonant S-B state exchange ($\sim \gamma_{SB}/\dot{\lambda}$) exceeds the relaxation time of the bath ($\sim 1/\omega_B \gamma_{BB}$). 
This predicts that the FF curves peel off from the UA one at $\dot{\lambda}/[\omega_B\,\cu^2] \approx \,\gu/\cu$, in good agreement with Fig.~\ref{fig:isothermality}.
We note that the speed regime where the FF protocol is effective {\em cannot} be treated in the Markovian approximation because the bath is not in local equilibrium on the time-scale of the ramp.
 
Fig.~\ref{fig:isothermality} also shows that the final temperature of S under FF driving deviates from $T$ at fast speeds. The FF protocol only thermalizes S at $\lambda(t)=0$. For $\lambda(t) \neq 0$, S is effectively decoupled from the bath and evolves adiabatically. Since the occupation number of S is fixed in an adiabatic process, its average energy increases as $\langle H(\lambda) \rangle = \sqrt{(1+\lambda)}\, T$ for $\lambda>0$. This gives $\langle H(\lambda_f) \rangle/T \approx 1 + \gu $ at small $\gu$ for $\dot{\lambda} \to \infty$, in quantitative agreement with Fig.~\ref{fig:isothermality}.

\section{Application: Heat Engine}

\noindent
The FF protocols can be used in the design of a highly efficient heat engine capable of producing a large power output. The engine uses the S oscillator as a working substance, with cold (C) and hot (H) reservoirs of optical phonons at temperatures $T_C$ and $T_H$ respectively. The model depicted in Fig.~\ref{fig:model}  describes the engine when S is coupled to C (H), with frequency $\omega_B=\omega_{C}$  ($\omega_H$) and detuning $\lambda(t) = (\omega_S^2(t) -\omega_B^2)/\omega_B^2$. Both reservoirs have the same coupling strengths $\gu$ and $\cu$. In a full cycle, S is first coupled to C and its frequency $\omega_S(t)$ is increased across resonance with $\omega_C$. Subsequently, S is coupled to H and $\omega_S(t)$ is decreased across resonance with $\omega_H$. 

\begin{figure}[htp]
\centering
  \includegraphics[width=1\columnwidth]{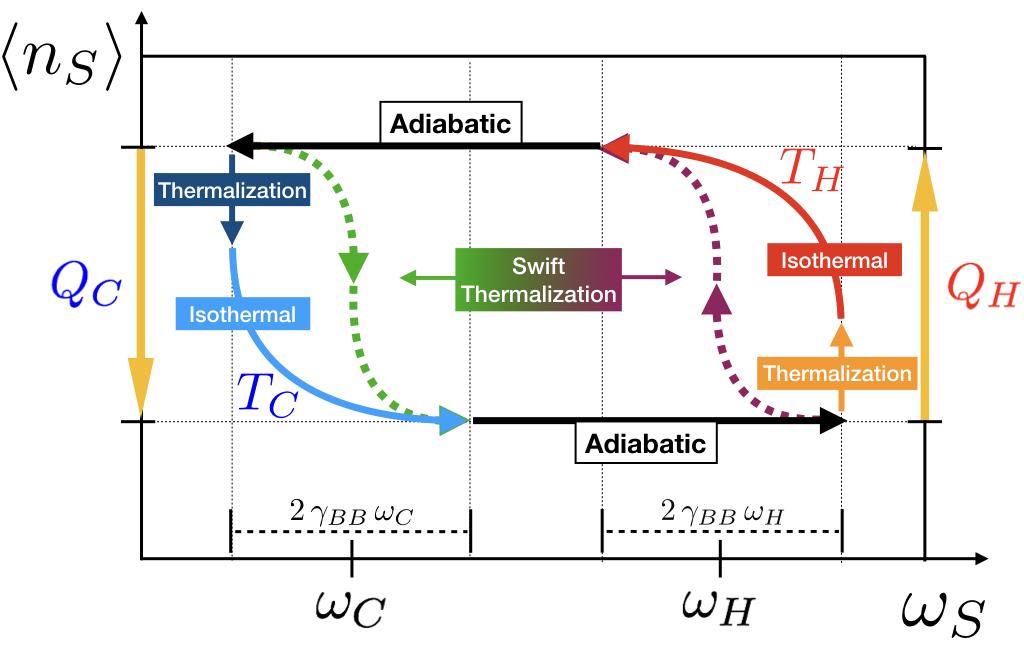}
    \caption{\textbf{Schematic $PV$ diagram for a heat engine:} The engine uses a harmonic working substance S and two reservoirs of optical phonons. During the thermal strokes, S draws heat $Q_H$ from a hot bath at $T_H$, and deposits heat $Q_C$ into a cold bath at $T_C$. During the adiabatic strokes (solid black curves), S is effectively isolated and $\langle {n}_S \rangle$ is constant. Solid colored curves show thermal strokes with slow driving, while colored dotted curves show swift thermalization with FF driving.}
  \label{fig:Eff1}
\end{figure}

Engines with a harmonic working substance behave much like ideal gas engines \cite{Tu,Deng,delCampo1,Kosloff2,Abah,Dechant, Arnaud}. For instance, one can define an effective pressure $P=\langle {n}_S \rangle$ and volume $V = \omega_S$ and construct a $PV$ diagram as shown in Fig.~\ref{fig:Eff1}. 

At slow speeds, the engine undergoes two `adiabatic' and `thermal' strokes. Consider for definiteness the forward ramp ($\dot{\lambda}>0$) with S coupled to C. During an adiabatic stroke $|\lambda(t)|\gg\gu,\cu$, S doesn't exchange energy with C, and $\langle {n}_S \rangle$ remains constant. When $|\lambda(t)|\lesssim\gu,\cu$, S can exchange energy with the bath and undergo a thermal stroke. The thermal stroke consists of two processes: (i) a thermalization process where S is brought to temperature $T_C$, (ii) an isothermal process where S remains at $T_C$ as $\omega_S$ is tuned across the bath's bandwidth. During an isothermal process, $\langle n_S \rangle=T_C/\omega_S$. Fig.~\ref{fig:Eff1} shows the four strokes (solid curves) in a complete slow cycle: (1) a contractive ($\dot{\lambda}>0$) thermal stroke with C, (2) a contractive adiabatic stroke, (3) an expansive ($\dot{\lambda}<0$) thermal stroke with H, and (4) an expansive adiabatic stroke. This cycle is generally irreversible because the thermalization process in each thermal stroke is irreversible. The degree of irreversibility is controlled by the ratio $r\equiv (T_C/T_H)/(\omega_C/\omega_H) \leq 1$. When $r=1$, the thermalization process is eliminated and we recover a Carnot efficiency (c.f. Sec. E in SI and Ref.~\cite{Arnaud}).

The engine can be sped up by implementing a FF protocol. At high ramp speeds, the protocol preserves the adiabatic strokes, but changes each thermal stroke into a swift thermalization process at the corresponding resonance. This results in an approximate Otto cycle; see the dotted colored curves in Fig \ref{fig:Eff1}. Intermediate speeds (not shown) result in a mix of partial thermalization, isothermal, and swift thermalization processes.

Over a cycle, S absorbs heat from H, uses some of this energy to do work, and releases the remainder into C. For a thermal stroke with either bath, we define the heat as $Q = |\Delta \langle H_{bath} \rangle|$, the change in the bath's average energy between the start and end of the stroke. Note the convention $Q_{C,H}>0$. Such heat may have contributions from spontaneous energy transfer (e.g. thermal conduction), as well as induced energy transfer (FF swift thermalization). The work done by the engine is then $W=Q_H-Q_C$, the difference between the absorbed and released heats. In the following, we consider two performance measures of the engine: (i) efficiency, given by $\eta = W/Q_H$, and (ii) average power over a cycle time $\tau$, measured by $P=W/\tau$. 

In the slow limit ($\dot{\lambda}\to 0$) we have [SI Text]:
\be
\eta= \eta_c - \frac{T_C}{T_H} \frac{(1-r)^2}{r\,(1-r+2\,\gu)},\quad\quad P\to0.
\label{Eq:EtaSlow}
\ee
For $0\leq r\leq1$, $\eta$ is bounded by the Carnot efficiency $\eta_C = 1-T_C/T_H$. At $r=1$, $\eta=\eta_C$.

\begin{figure}[htp]
\centering
  \includegraphics[width=1.02\columnwidth]{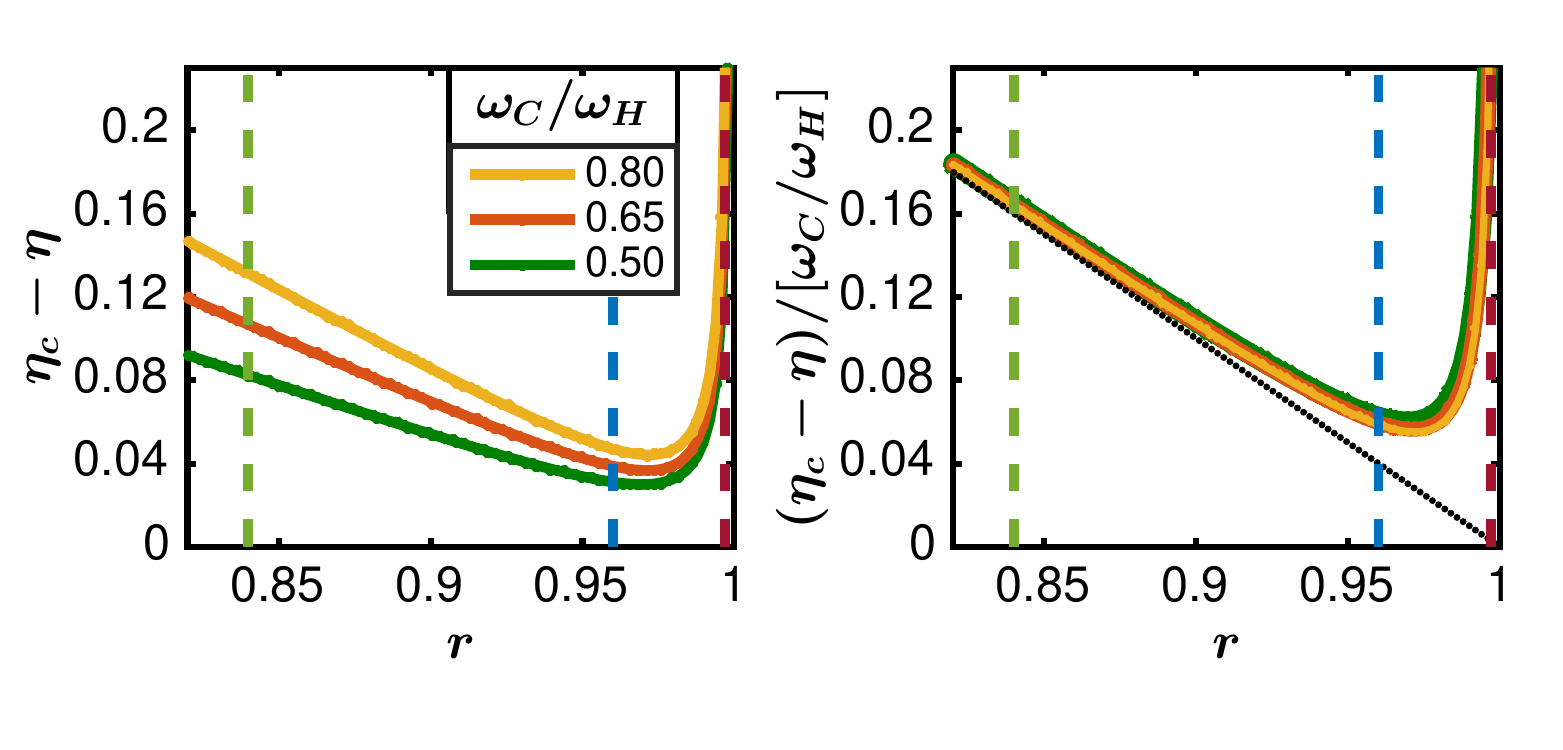}
    \caption{ \textbf{High-speed engine efficiency $\eta$ as a function of $r \equiv (T_C/T_H)/(\omega_C/\omega_H)$.} (Left) The difference between the Carnot efficiency $\eta_c$ and the efficiency $\eta$ for several values of $\omega_C/\omega_H$. $\eta_c-\eta$ is minimized at a special $r=r_{min}\approx 0.96$, and grows sharply as $r$ approaches the breakdown value $r_0\approx 1$. (Right) Upon re-scaling by $\omega_C/\omega_H$, the curves collapse onto each other. The dark line is the zero bandwidth limit $\gu=0$. The vertical dashed lines indicate the values of $r=0.84<r_{min}$ (green), $r =0.96\approx r_{min}$ (blue), and $r_{min}<r=0.997<r_0$ (scarlet) used in Fig.~\ref{fig:Eff3}. Simulation parameters: $N=100$, $\gu = 0.03$, $\cu = 0.02$, $T_H=100$. $T_C$ is varied to tune $r$. }
  \label{fig:Eff2}
\end{figure}

For fast enough ramps ($\cu,\,\gu \ll \dot{\lambda}/\omega_{C,H}$), we break up the analysis into two cases based on the relation between $\cu$ and $\gu$. 

When $\gu\ll\cu \ll1$, S effectively interacts with a single bath oscillator B of frequency $\omega_{H,C}$ during either thermal stroke. The FF protocol induces an exchange of thermal occupation distributions between S and B, so that 
\be
Q_{H,C} = \omega_{H,C} (\langle n_H \rangle-\langle n_C\rangle).
\ee
Above, the H and C baths are taken to be in the classical regime, so that $\langle n_{H,C}\rangle \sim T_{H,C}/\omega_{H,C}$. Then the efficiency and power are given by 
\be
\eta= 1-\frac{\omega_C}{\omega_H} \leq \eta_{c},\quad\quad P=\frac{k_B T_H}{\tau}\,\eta\, \left(1-r\right)
\ee
This efficiency is characteristic of an Otto engine. It is bounded by the Carnot efficiency, as follows from the consistency condition $\langle n_H\rangle \geq \langle n_C\rangle$. While it is possible to attain the Carnot efficiency in the limit $r\to1$, the power output simultaneously tends to zero. To achieve finite power in practice, one must keep $r<1$ at the expense of some efficiency. We note that one can also optimize $P$ with respect to the ratio $\omega_C/\omega_H$ at fixed $T_C,\, T_H$; the corresponding efficiency is then the well-known Curzon-Ahlborn bound~\cite{Abah,Dechant}.

When $\gu\gtrsim\cu$, the finite bandwidth of the bath modifies the heat at high-speeds
\be\label{Qff}
Q_{i} = \omega_{i}\bigg(\frac{T_H}{\omega_H} - \frac{T_C}{\omega_C}\bigg)(1+\gu^2) \mp 2\, T_{i}\, \gu^2.
\ee 
We take $i=H$ and the negative sign for the heat released by the hot reservoir, and $i=C$ and the plus sign for the heat released into the cold reservoir. The correction $\mathcal{O}(\gu^2)$ arises because FF is no longer transitionless. It induces excess excitations in the bath (i.e. dissipation) during the S-B exchange. This causes S to extract less net heat from H and dump more into C, reducing the efficiency [SI Text]. 

Fig.~\ref{fig:Eff2} shows the high-speed efficiency for $\gu\sim\cu$ as a function of $r$ for several ratios $\omega_C/\omega_H$. As $r$ is increased, the difference between $\eta$ and $\eta_c$ decreases until a minimum is reached at a specific value $r_{min} = 1 - \mathcal{O}(\gu)$. By tuning $r$ close to $r_{min}$, the engine can operate near the Carnot efficiency. If we continue to increase $r > r_{min}$, then $\eta$ diverges from $\eta_c$, and the engine eventually breaks down. The breakdown value $r_0 = 1-\mathcal{O}(\gu^2)$ occurs when $Q_H = Q_C$ and the engine fails to extract useful work.  The figure also shows a collapse of curves upon re-scaling by $\omega_C/\omega_H$. Thus $(\eta-\eta_C) \propto \omega_C/\omega_H$ can be taken arbitrarily close to zero by decreasing $\omega_C/\omega_H$ to further optimize the efficiency. For details, see [SI Text].

\begin{figure}[htp]
\centering
  \includegraphics[width=1.03\columnwidth]{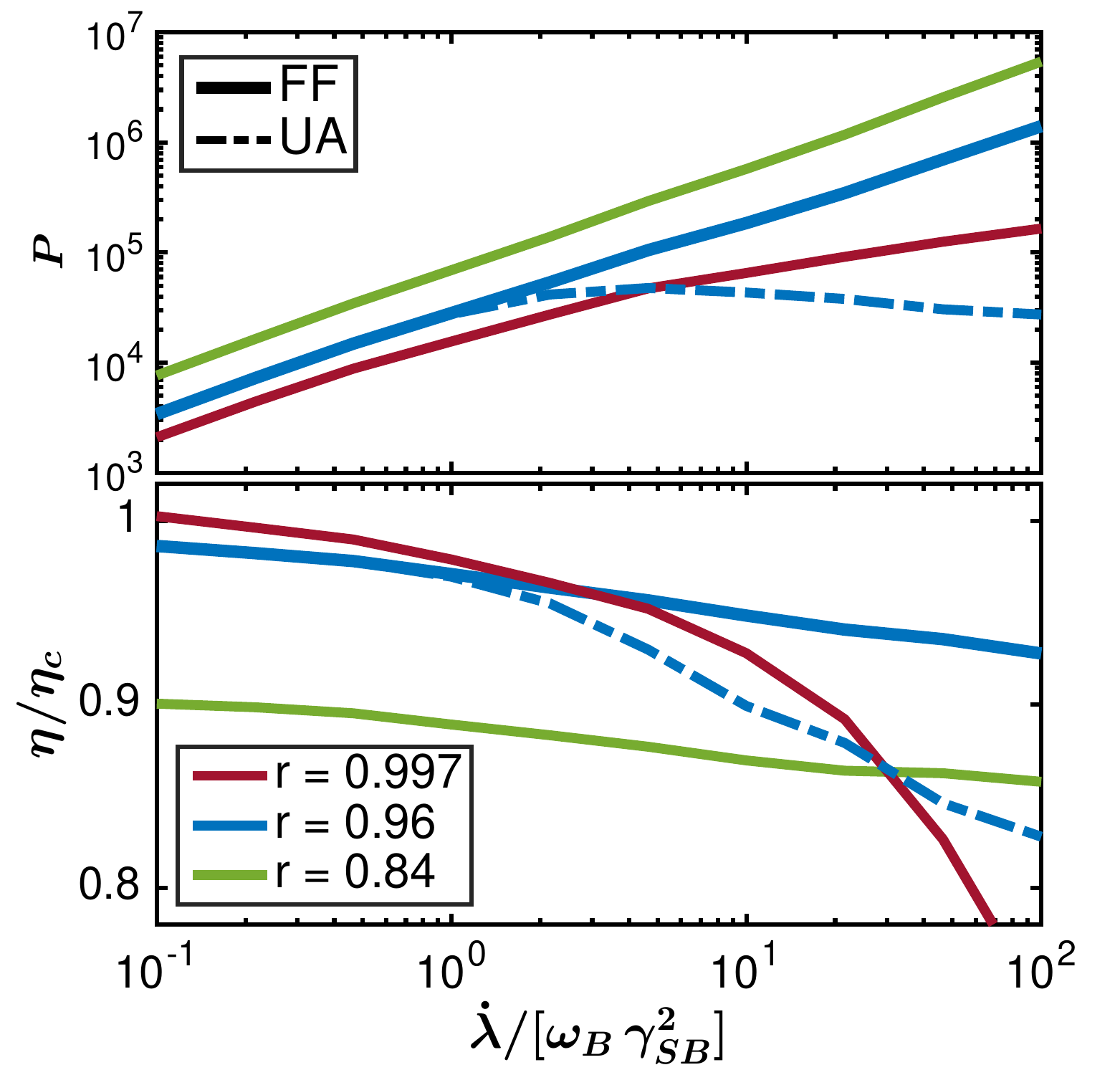}
    \caption{ \textbf{FF driving produces an efficient high-power engine.} (Top panel) Plot of the engine's power output $P$ as a function of ramp speed $\dot{\lambda}/[\omega_B \cu^2]$. Power decreases with $r$. At $r=0.96$, FF outperforms UA in producing power at high-speeds. (Bottom panel) Plot of the relative efficiency $\eta/\eta_c$ as a function of ramp speed. $\eta/\eta_c$ first increases with r (curves $r=0.84,0.96$), then shows signs of breakdown at high speeds, when $r$ approaches 1 (curve $r=0.997$). At $r=0.96$, FF beats the UA efficiency at high-speeds. Simulation parameters: $N=100$, $\omega_C=1$, $\omega_H=2$, $\gu = 0.03$, $\cu = 0.02$, $T_H=100$. $r$ is varied using $T_C$.}
  \label{fig:Eff3}
\end{figure}

The engine's performance in the regime $\gu\gtrsim\cu$ across several speed scales is summarized in Fig.~\ref{fig:Eff3}. The plot shows power $P$ and relative efficiency $\eta/\eta_c$ for FF driving (solid curves) with three different values of $r$: $r<r_{min}$, $r\approx r_{min}$, and $r_{min} < r < r_0$ (see vertical dashed lines in Fig.~\ref{fig:Eff2}), and for UA driving (dashed curves) at $r \approx r_{min} = 0.96$. In the FF protocols, the power decreases with increasing $r$ over the whole speed domain (recall $P\to 0$ as $r \to 1$). At a given $r$, $P$ increases with ramp speed. As $\dot{\lambda} \to \infty$, $P$ becomes linearly proportional to $\dot{\lambda}$, since the work done by the engine approaches a constant value. The power over a cycle in the UA protocol is not only lower than the corresponding FF power at large speeds, but also decreases with $\dot{\lambda}$. The bottom panel shows that the relative efficiency of FF protocols increase with $r$ for $r<r_{min}$ over the whole speed domain. For $r>r_{min}$, the $r=0.997$ curve shows signs of the engine breakdown at high speeds. At a given $r$, $\eta/\eta_c$ decreases from its slow speed value in Eq.~\eqref{Eq:EtaSlow} to its fast speed value derived from Eq.~\eqref{Qff}, see SI Text. The efficiency of the UA protocol is less than corresponding FF protocol at large speeds.
Thus Fig.~\ref{fig:Eff3} establishes that FF outperforms UA in both power and efficiency.  

Since $P\sim\tau^{-1}\sim \dot{\lambda}$ with FF protocols, $P$ can in principle be arbitrarily enhanced by reducing the cycle time $\tau$. There is, however, a practical limit to how small we can make $\tau$ while running the engine without interruption. Any fast cycle takes the oscillator B out of equilibrium due to the S-B exchange induced by FF. Thus B must be given enough time ($\sim (\omega_{B}\gu)^{-1}$) to equilibrate with the remaining bath degrees of freedom before the next cycle. This imposes a ramp speed bound $\dot\lambda\lesssim 2\,\omega_B\,|\lambda_f - \lambda_i|\, \gu$, which limits the maximum power output. The simulation results presented here satisfy this condition. One can overcome this constraint and further increase the power output by reconnecting S to different parts of the bath after each cycle.

\section{Methods}
\noindent
\textbf{RW-FF protocol.} Consider the CD Hamiltonian for two coupled oscillators in equations ($\ref{H_CD}$) and (\ref{Gauge_pot}). A rotation wave (RW) approximation is obtained by writing $H_{CD}$ in terms of the creation and annihilation operators
\begin{eqnarray}\nonumber
a_S&\equiv&\sqrt{\frac{\omega_B}{2}} \left( X+i \frac{P}{\omega_B}\right),\,\,\, a_B\equiv \sqrt{\frac{\omega_B}{2}} \left(x_B+i\frac{p_B}{\omega_B}\right)
\end{eqnarray}
and keeping only number conserving terms to obtain
\begin{multline}
H_{CD}\approx \omega_B \bigg[\bigg(1+\frac{\lambda}{2}\,\bigg) a_S^{\dagger} a_S+ a_B^{\dagger} a_B-{\frac{\cu}{2}} (a_S^{\dagger} a_B+a_B^{\dagger} a_S)\\ \label{HCDRW}
+\frac{\dot\lambda}{i\,\omega_B} \frac{\cu}{ (\lambda^2+4\cu^2)} (a_S^{\dagger} a_B-a_B^{\dagger} a_S )\bigg].
\end{multline}
A few comments are in order. First, we define $a_S$ using $\omega_B$ instead of $\omega_S(t)$ to avoid introducing additional time-dependent corrections into the Hamiltonian. This construction is adequate since the dominant effects occur near resonance. Next, we have omitted an additive constant which has no effect on dynamics. Finally, $H_{CD}$ in Eq.~\eqref{HCDRW} has the same form as in the Landau-Zener (LZ) two-level problem [SI Text]. 

We obtain the rotating-wave FF protocol from Eq.~\eqref{HCDRW} by the simple rotation
\be\label{RW Rotation}
a_S\to a_S\, \mathrm e^{i\theta_S},\quad \tan(\theta_S)=\frac{\dot \lambda}{\omega_B (\lambda^2+4\,\cu^2)}.
\ee
The corresponding unitary $U= e^{i \theta_S a_S^\dagger a_S}$ which generates the $H_{FF}^{RW}$ from $H_{CD}$ is analogous to a unitary previously obtained for the LZ problem~\cite{Bukov1}. It automatically satisfies the boundary conditions $U(t_i)=U(t_f)=I$ if $\dot\lambda$ vanishes at the protocol boundaries.

In the original phase space variables, we obtain:
\be 
H_{FF}^{RW}=\frac{P^2}{2} + \frac{\tilde{\omega}_{S}^2(t)}{2} X^2 +  \frac{p_B^2}{2} + \frac{\omega_{B}^2}{2} x_B^2 -  \tcu(t)\omega_B^2\,x_B X,
\ee
where $\tilde{\omega}_{S}^2(t)$ and $\tcu(t)$ are given by Eq.~\eqref{RW2} and \eqref{RW3}, respectively.\\

\noindent
\textbf{FE-FF protocol.}
A Hamiltonian can obtained from $H_{CD}$ by a sequence of unitary transformations which mix the degrees of freedom of both S and B (see SI Text):
$$ H_{FF}^{'} = \frac{P^2}{2} + \frac{\Lambda'(t)}{2} X^2 + \frac{p_B^2}{2} + \frac{K'(t)}{2} x_B^2 - C'(t)\,X\,x_B $$
where now ${\Lambda}'(t),\,K'(t), \, C'(t)$ are non-trivial functions of time. The need to modulate $K'(t)$ makes $H_{FF}^{'}$ not an experimentally viable protocol because it requires additional control of an inaccessible bath parameter. 

To realize $H_{FF}^{'}$, we construct a Floquet-Engineered Hamiltonian:
\be 
H_{FF}^{FE}=\frac{P^2}{2} + \frac{\tilde{\omega}_{S}^2(t)}{2} X^2 +  \frac{p_B^2}{2} + \frac{\omega_{B}^2}{2} x_B^2 -  \tcu(t)\omega_B^2\,x_B X,
\ee
where
\begin{align} \label{FE1}
\tilde{\omega}_{S}^2(t)\, &=  {\Lambda}'(t)+\omega_B^2 - K'(t)\\
\tcu(t)\,\omega_B^2\, &= C'(t) - \sqrt{2(K'(t) - \omega_B^2)}\,\, \Omega \,\cos(\Omega \,t) \label{FE2}
\end{align}
Fig.~\ref{fig:fidelity}b illustrates Eqns.~(\ref{FE1}) and (\ref{FE2}).  When $\Omega^{-1}$ is the smallest timescale in the problem, the dynamics under $H^{FE}_{FF}$ can be treated perturbatively in $1/\Omega$ using a high-frequency Magnus expansion~\cite{Bukov2}. To leading order, the effective Floquet Hamiltonian coincides with $H_{FF}^{'}$. In the SI Text, we detail the stroboscopic equivalence of $H^{FE}_{FF}$ and $H_{FF}^{'}$ and show that $H^{FE}_{FF}$ is a FF protocol which implements the complete CD protocol as $\Omega\to\infty$.\\

\section{Discussion and Conclusion}\label{sec8}

\noindent
We have developed efficient FF protocols which realize a resonant state exchange between a system and a bath oscillator by controlling the local parameters of the system and the system-bath coupling. In the presence of a phonon bath, these FF drives realize a swift thermalization process with high fidelity. We used these FF protocols in the design of a high-power engine which can operate near the Carnot efficiency. Our work demonstrates the power of FF methods to achieve efficient energy transfer in small open quantum systems and optimize thermodynamic processes. With recent advances in reservoir engineering~\cite{Koch}, this opens up the possibility of realizing powerful efficient microscopic engines with non-Markovian environments. 

Interestingly, the FF protocols are most efficient at fast driving speeds, where the bath does not relax and {\em cannot} be treated in the Markovian approximation. The FF protocols realize a coherent exchange of energy with a local bath degree of freedom, which subsequently relaxes with the rest of the bath. In the limit of zero bath-bath coupling (and hence infinite bath relaxation time), the local bath degree of freedom does not relax after the exchange, resulting in no irreversible energy dissipation. At finite bath-bath coupling, a small amount of residual energy is dissipated due the mismatch of the final state of the local bath degree of freedom and its equilibrium state. This mistake is controlled by the bandwidth of the bath and is independent of the protocol ramp speed (c.f. Eq.~\eqref{Qff}). Thus our protocols are different from those previously obtained with Markovian environments~\cite{Martinez, Chupeau, Dann, Li,Patra,Boyd}, where quick equilibration is achieved at the expense of dissipative losses that increase with the ramp speed.

The approach presented in this article applies broadly to systems with Landau-Zener characteristics, where adiabatic state exchanges occur as a consequence of avoided level crossings. In such setups, FF driving can be used for rapid state preparation. Using swift thermalization, one can cool many-body quantum systems close to their ground state, of interest in numerous applications of ultra-cold atom and optomechanical systems.  \\ 

\noindent\textbf{ACKNOWLEDGEMENTS.} We thank Dries Sels and Chris Laumann for useful discussions. We are pleased to acknowledge that the computational work reported on in this paper was performed on the Shared Computing Cluster which is administered by Boston University’s Research Computing Services. This work was supported by AFOSR FA9550-16-1-0334 (A.P.), NSF DMR-1813499 (T.V. and A.P.), and NSF DMR-1752759 (T.V. and A.C.). A.C. acknowledges support from the Sloan Foundation through Sloan Research Fellowships.

\newpage
\afterpage{\blankpage}
\afterpage{\blankpage}
\newpage

\section{Supplemental Information}

\subsection{Appendix A: Two oscillator system}\label{sec: AppA}
\noindent
\textbf{Hamiltonian.} The system S consists of a particle $(X,P)$ in a tunable harmonic potential, which is locally coupled to a bath oscillator B with coordinates $(x_B,p_B)$. The Hamiltonian is:
\begin{equation}\label{H0}
H_0 = \frac{1}{2}P^2 + \frac{\tilde{\omega}_{S}^2(t)}{2} X^2  -\tcu(t) \,\omega_B^2\, x_{B}\,X +  \frac{1}{2}p_{B}^2 +  \frac{\omega_{B}^2}{2} x^2_{B} 
\end{equation}
where $\tilde{\omega}_S(t)$ is the system's time-dependent frequency, $\omega_B$ is the frequency of the bath mode, and $\tcu(t)$ is the dimensionless S-B coupling. Unassisted (UA) protocols set a target ramp $\tilde{\omega}_S(t) = \omega_S(t)$, which fast-forward (FF) protocols modify to achieve fast adiabatic driving. In unassisted (UA) protocols, $\tcu(t) = \cu \ll 1$ is held constant during the ramp, while in fast-forward (FF) protocols, $\tcu(t)$ is enhanced in time near resonance. In all cases, the value of the S-B coupling at the start and end of the ramp is given by $\cu$: $\tcu(t_i)=\tcu(t_f)=\cu$. 
\\

\noindent
\textbf{Normal Mode Dispersion.} In terms of the normal mode occupation number operators $n_{+}$ and $n_{-}$, $H_0$ becomes:
\begin{equation}
H_0 = \omega_{+}(t) \big(n_{+} +1/2\big) + \omega_{-}(t) \big(n_{-}+1/2\big)
\end{equation}
where
\begin{equation}\label{dispersion}
\omega^2_{\pm}(t) = \omega_{B}^2 \,\bigg(1 + \frac{\lambda(t)}{2} \pm \sqrt{ \bigg[\frac{\lambda(t)}{2}\bigg]^2 + \cu^2 }\, \bigg)
\end{equation}
and
\begin{equation}
\lambda(t)\equiv\frac{\omega_S^2(t)-\omega_B^2}{ \omega_B^2}.
\end{equation}
Here, $\lambda(t)$ measures the detuning of an UA drive from resonance $\omega_S(t) = \omega_B$.
The dispersion in equation (\ref{dispersion}) is shown in Figure 2 of the main text. \\
 
\subsection{Appendix B: Emergent speed scales in unassisted protocols}\label{sec: AppB}
 
 \noindent
\textbf{Emergent Speed Scales.} In unassisted protocols, the response of the S+B system depends on how the ramp speed $\dot{\lambda}$ compares to two emergent speed scales. We present a heuristic derivation of these scales.

Consider a transition from the energy level with $(n,m)$ quanta in the $(+,-)$ normal modes to the energy level with $(n',m')$ quanta. The energy change is:
\begin{equation}
\Delta(t) \equiv E_{n,m} - E_{n' , m'} = \delta n \,\, \omega_{+}(t) + \delta m \,\, \omega_{-}(t) 
\end{equation}
where $\delta n = n - n' $ and $\delta m = m - m' $. Such a transition can be classified based on the relation between $\delta n$ and $\delta m$: (i) $\delta n = - \delta m$ for an exchange process which conserves the total number of quanta, (ii) $\delta n = \delta m$ for a pair creation/annihilation process between the normal modes, and (iii) $\delta n \neq \pm \delta m$ for processes that create/destroy quanta within each normal mode. 

A transition process has negligible probability of occurrence when the gap $\Delta(t)$ is varying slowly enough:  
\[
\bigg|\frac{d \Delta(t)}{d t} \bigg| \frac{1}{\Delta(t)} \ll \Delta(t)\quad \Leftrightarrow \quad |\dot\lambda|\ll \frac{\Delta^2(\lambda)}{|\partial_\lambda \Delta(\lambda)|}.
\]
At a given $\dot{\lambda}$, any transition process satisfying this condition is considered inactive and essentially adiabatic. The energy gap reaches its minimum value at resonance $|\lambda|\approx\cu$. For the UA protocol, the adiabatic condition is first violated near resonance at the speed scale:
\be
\dot{\lambda}_1 \equiv \frac{\Delta^2}{|\partial_\lambda \Delta| }\bigg|_{|\lambda| \sim \cu} \sim  \omega_B\,\cu^2
\ee
When $\dot\lambda$ becomes comparable or larger than $\dot \lambda_1$, number-conserving non-adiabatic transitions satisfying $\delta n=-\delta m$ start to occur.

At even faster speeds, when $\dot\lambda$ becomes comparable to
\be
\dot{\lambda}_2 \equiv \frac{\Delta^2(\lambda)}{|\partial_\lambda \Delta(\lambda)|}\bigg|_{|\lambda| \sim \cu} \sim \omega_B,
\ee
the pair creation/annihilation processes with $\delta n+\delta m\neq 0$ occur. These processes lead to the breakdown of the rotating wave (RW) approximation used to develop a simple FF protocol in the main text. Since $\cu\ll 1$, $\dot{\lambda}_1 \ll \dot{\lambda}_2$. Therefore there is a large window of protocol speeds where one can rely on the rotating wave approximation and use the simplified RW-FF protocol.\\

\noindent
\textbf{Number-Conserving Regime.} When the condition $ \dot{\lambda} \ll \dot{\lambda}_2$ is satisfied, there is a mapping of $H_0$ to the Landau Zener (LZ) problem. To see this, express $H_0$ in terms of creation/annihilation operators and drop all number non-conserving terms: 
\begin{equation} \nonumber
H_0 \approx  \begin{bmatrix}
   a^{\dagger} &  b^{\dagger} \\
\end{bmatrix}   
 \begin{bmatrix}
    \omega_S & - g\,\, \\
    -g\quad  & \omega_B \\
\end{bmatrix}
 \begin{bmatrix}
   a \\  b \\
\end{bmatrix} 
\end{equation}
where $g \approx \frac{1}{2}\,\omega_B\, \cu$ near resonance, $(a^{\dagger},a)$ are the system bosonic creation/annihilation operators, and $(b^{\dagger},b)$ are the creation/annihilation operators of oscillator B. 

Interpreting $a$ and $b$ as Schwinger bosons, we write the Hamiltonian using the angular momentum operators~\cite{Auerbach}
\begin{multline}
H_0 =(\omega_S -\omega_B)\, L_z  - 2g \,L_x +  \frac{(\omega_S+\omega_B)}{2}N_b \,,\\
L_z=\frac{1}{2}(a^\dagger a-b^\dagger b),\; L_x=\frac{1}{2}(a^\dagger b+b^\dagger a),\;N_b=a^\dagger a+b^\dagger b.
\label{eq:Hamiltonian_LZ}
\end{multline}
The total number of bosons $N_b$ is conserved and sets the total angular momentum of the system $L=N_b/2$. When $N_b=1$ and hence $L=1/2$, this Hamiltonian is equivalent to the LZ Hamiltonian with gap $g$ and tuning parameter $\lambda_{LZ}(t) \equiv \frac{1}{2}(\omega_S(t) -\omega_B)$. Because the Hamiltonian~\eqref{eq:Hamiltonian_LZ} is linear in the angular momentum operators, the solution in the Heisenberg picture is independent of $L$ or $N_b$. Therefore, one can use well-known results of the LZ problem for identifying the adiabatic breakdown criterion for general $L$ and for finding CD and FF protocols. In particular, the characteristic LZ ramp speed defining the adiabatic-diabatic crossover is $\dot{\lambda}_{LZ} \sim g^2$~\cite{AP1,Shevchenko}, which is equivalent to $\dot{\lambda}_1 \sim \omega_B\,\cu^2$ for the corresponding oscillator problem.

\begin{figure}[htp]
\centering
  \includegraphics[width=1.06\columnwidth]{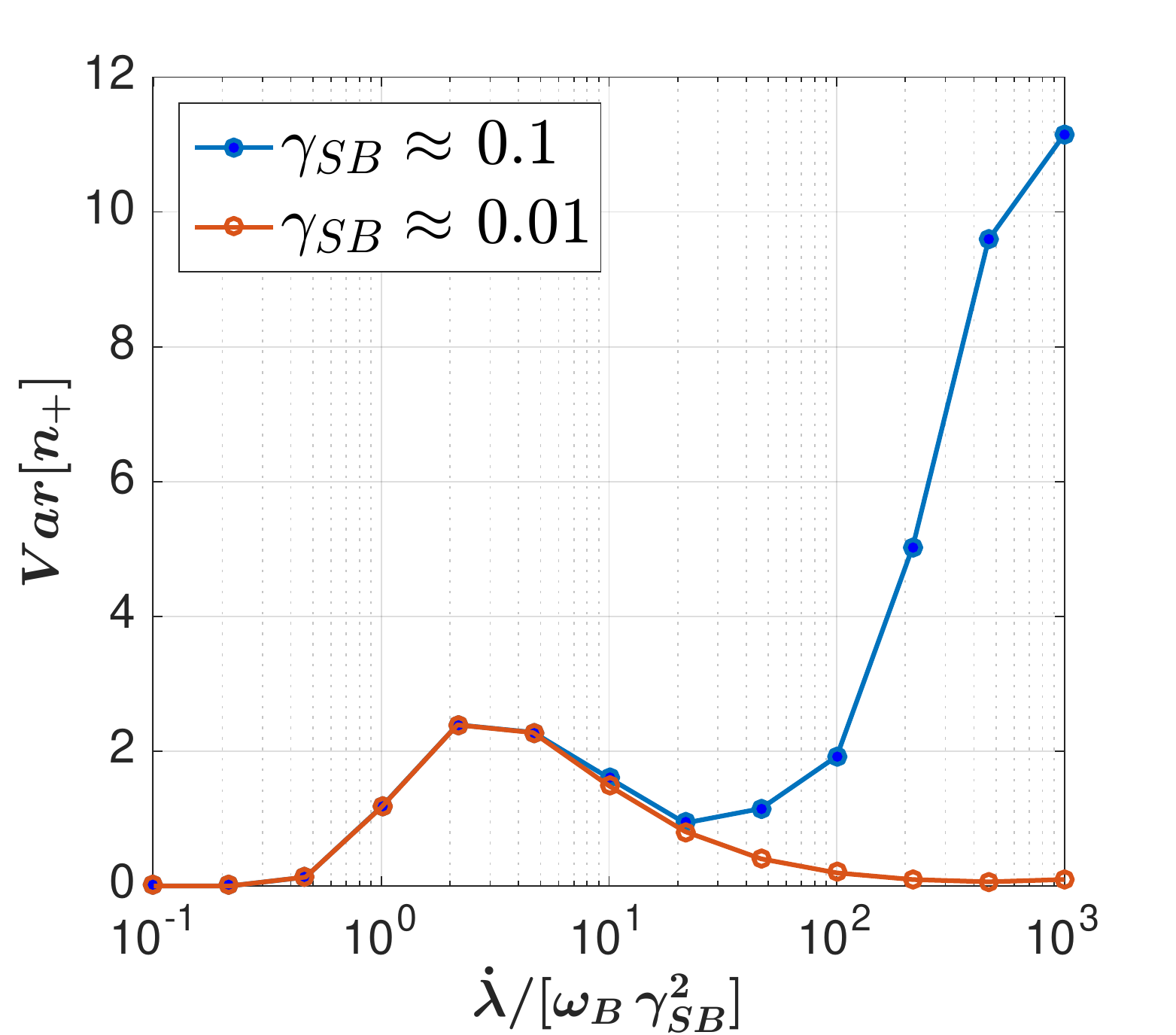}
    \caption{\textbf{Number variance of the $(+)$ normal mode versus normalized speed after an unassisted ramp.} The two curves correspond to two different values of the system-bath coupling. There is a good collapse of the results in the number-conserving regime $|\dot\lambda|\ll \omega_B$, where the dynamics is equivalent to that of the LZ problem. The collapse breaks down at higher speeds, where the second scale $\dot{\lambda}_2 \sim\omega_B$ becomes relevant. The system is initialized in an eigenstate $| n_{-},n_{+}\rangle = |3,1\rangle$. Variance is measured at the end of the ramp. Simulation Parameters: $\lambda_i = -0.8$, $\lambda_f = +0.8$, and $\omega_B = \sqrt{5}$.}
  \label{fig:CollapseScales}
\end{figure}

In this number conserving or LZ regime, the ramp speed scale $\dot{\lambda}_1$ dominates the physical behavior of the system. Thus physical quantities show a collapse of curves when re-scaling $\dot{\lambda}$ by $\dot{\lambda}_1$. As an example, Fig.~\ref{fig:CollapseScales} shows the occupation number variance of the (+) normal mode after an unassisted ramp $\lambda(t)$ across resonance, as a function of $\dot{\lambda}/\omega_B \, \cu^2$. The plot shows a good collapse of curves over different values of $\cu \approx 0.1, 0.01$ in the regime $\dot\lambda\ll \dot \lambda_2$.

\subsection{Appendix C: Two-Particle Counter-diabatic Drive }\label{sec: AppC}
\noindent
\textbf{Counter-diabatic Gauge Potential.} For any protocol $\lambda(t)$, one can design dynamics which follow the instantaneous eigenstates of $H_0[\lambda(t)]$ in accordance with the adiabatic theorem. This is accomplished by evolving the system under the counter-diabatic Hamiltonian $H_{CD} = H_0 + \dot{\lambda} \mathcal{A}$, where the gauge potential $\mathcal{A}$ satisfies the commutator relation~\cite{Kolodrubetz,Dries}:
\begin{equation}
[H_0, i\,\hbar\,\partial_{\lambda} H_0 + [H_0,\mathcal{A}]] = 0.    
\end{equation}

For $H_0$ in equation (\ref{H0}), we find
\be
\mathcal{A} = a_{1} \frac{\{X\,, P\}_{+}}{2} + a_{2} \,X \,p_B +a_{3}\, x_B \,P + a_{4} \frac{\{x_B\,, p_B\}_{+}}{2} 
\ee
where 
\begin{align}
a_{1} &= - \frac{\lambda^2 + \cu^2\,(2-\lambda)}{4(1+\lambda-\cu^2)(\lambda^2+ 4\,\cu^2)}  \\
a_{2} & = +\frac{\cu\,(4(1+\lambda)+\lambda-6\cu^2)}{4(1+\lambda-\cu^2)(\lambda^2+ 4\,\cu^2)} \\  
a_{3} & = -\frac{\cu\,(4(1+\lambda)-\lambda-2\cu^2)}{4(1+\lambda-\cu^2)(\lambda^2+ 4\,\cu^2)}  \\  
a_{4} & = - \frac{\cu^2(2+\lambda)}{4(1+\lambda-\cu^2)(\lambda^2+ 4\,\cu^2)}
\end{align}

In the weak coupling regime $\cu \ll 1$ and close to the resonance $|\lambda|\ll 1$, these expressions simplify:
\begin{align}\label{WeakC1}
a_{1} &\approx - \frac{K_1}{4(1+\lambda)} \\ \label{WeakC2} 
a_{2} &\approx -a_3\approx \frac{\cu}{(\lambda^2+ 4\,\cu^2)} \\  \label{WeakC3}
a_{4} &\approx  - \frac{1}{4(1+\lambda)} \frac{2\,\cu^2}{(\lambda^2+ 4\,\cu^2)}
\end{align}
The factor $K_1$ in Eq.~\eqref{WeakC1} is close to $1/2$ near resonance ($|\lambda| \ll \cu$) and smoothly approaches $1$ as $\lambda \to \pm \infty$. Note that $a_2$ and $a_3$ are much larger than $a_1$ for $|\lambda| \ll \cu$; we therefore set $K_1 = 1$ with negligible error. Moreover, $a_4 \sim \cu \, a_2$ so it can be ignored to leading order. The expressions in Eqs.~\eqref{WeakC1} and \eqref{WeakC2} appear in the main text in Eq.~(5).\\

\noindent
\textbf{Dynamic Switch under $H_{CD}$.} The dynamics under $H_{CD}$ is most simply seen in the $\cu\to0$ limit, in which $ a_1 \to -1/[4(1+\lambda)]$, $ a_2=-a_3 \to (\pi/2) \delta(\lambda)$, and $a_4 \to 0$. When $\lambda \neq 0$, $H_{CD}$ reduces to the well-known result for a dilated oscillator in vacuum~\cite{Kolodrubetz,Deffner} $$H^{(0)}_{CD} = -\frac{\dot{\lambda}}{4(1+\lambda)}\big(X \,P + P \,X\big).$$ Near resonance $\lambda \approx 0$, the equations of motion become 
\begin{align}
\dot X &\approx - \frac{\pi}{2}\, \delta(t - t_c) \, x_B, \quad\quad\quad \dot P &\approx - \frac{\pi}{2}\, \delta(t - t_c) \, p_B  \\
\dot x_B &\approx+ \frac{\pi}{2}\, \delta(t - t_c) \, X, \quad\quad\quad\dot p_B &\approx+ \frac{\pi}{2}\, \delta(t - t_c) \, P 
\end{align}
where $t_c$ is the time at which the system is at resonance, i.e. $\lambda(t_c)=0$. Solving these equations in the time interval $[t^-=t_c-\epsilon,t^+=t_c+\epsilon]$, with infinitesimal $\epsilon > 0$, we find $$X(t^{+}) = - x_B(t^{-}) \,,\, x_B(t^{+}) =  X(t^{-})$$ $$P(t^{+}) = - p_B(t^{-}) \,,\, p_B(t^{+}) =  P(t^{-})$$ 
Up to a minus sign, the counter-diabatic protocol forces a swap of the phase space coordinates $(X,P)$ of the system particle with those of the bath mode $(x_B,p_B)$. As the character of the normal modes change from S to B and vice-versa across resonance, the swap ensures the preservation of the occupations of the normal modes of $H_0$ across resonance. Before and after the swap, the occupation numbers are preserved by driving the system with $H^{(0)}_{CD}$.

\subsection{Appendix D: Fast Forward Drive }\label{sec: AppD}

In this section, we derive a FF Hamiltonian which implements $H_{CD}$ with accessible controls using Floquet engineering. The task is achieved in two steps: (i) We transform $H_{CD}$ using a series of unitary rotations $U_k$ to obtain a
fast-forward Hamiltonian $H_{FF}'$ with three time-dependent couplings: $\tilde{\omega}_S^2(t),\,\tcu(t),\, \tilde{\omega}_B^2(t)$. (ii) In order to eliminate the time dependence in the bath frequency $\tilde{\omega}_B^2(t) \to \omega_B^2$, we apply an additional periodic modulation of the system-bath coupling $\tcu(t)$ to generate a Floquet-Engineered FF Hamiltonian equal to $H_{FF}'$ in the limit of high driving frequency.\\

\noindent
\textbf{(i) Unitary transformations.} We shall construct a sequence of four time-dependent unitary transformation $U_{k}(t)$, $k=1,\dots 4$ yielding Hamiltonians $H_{k}$ equivalent to $H_{CD}$: $$H_k= U_{k}^{\dagger} H_{k-1} U_{k} - i U_{k}^{\dagger} \partial_t U_{k},\,\quad\quad H_{k=0} \equiv H_{CD}. $$

Each unitary will depend explicitly only on $\lambda$ and its time derivatives up to order 5. $\lambda(t)$ is chosen sufficiently smoothly such that $U_k(t)=I$ and $\partial_t U_k(t)=0$ at the beginning and the end of the protocol. To do this, we impose that time the derivatives $\lambda^{(j)}$, $j\leq6$, vanish at the ramp boundaries. 

The condition $U_k(t_{i,f})=I$ ensures that the FF protocol retrieves the target adiabatic state at the end of the ramp. To see this, consider the $n$-th eigenstate of $H_0$ evolved under $H_{CD}$: $|\psi_{CD}(t)\rangle=|\psi_n(\lambda(t))\rangle$. The wave function under time evolution with the rotated Hamiltonian $H_4$ follows this eigenstate rotated by the corresponding unitary~\cite{Kolodrubetz}
\[
|\psi_4(t)\rangle=U_4(t) U_3(t) U_2(t) U_1(t) |\psi_{CD}(t)\rangle.
\]  
Since each unitary is the identity at the protocol boundaries, $|\psi_4(t)\rangle$ coincides with the target $|\psi_n(\lambda(t))\rangle$ at the beginning and end of the ramp.

The condition $\partial_t U_k(t_{i,f})=0$ ensures $H_4 = H_0$ at the protocol boundaries. This requirement guarantees the stability of the final state after the ramp. Otherwise, any target eigenstate of $H_0$ would not be an eigenstate of $H_4$, and would not be stationary after the ramp (see e.g. Ref.~\cite{Ness}).

The unitaries $U_k$ are designed to successively eliminate momentum-dependent couplings. The first two unitaries,
$$ U_{1} = \exp\bigg( - i \bigg[ (\eta+ \dot{\lambda} a_1)\, \frac{X^2}{2} + \dot{\lambda}\, a_2 \,X\,x_B+ \dot{\lambda} \,a_4 \,\frac{x_B^2}{2} \bigg]\bigg) $$ and $$ U_{2} = \exp\bigg( i \,\mu\frac{P^2}{2}\bigg), $$
where 
$$\eta(t) \equiv (\dot{\lambda}\, \dot{a}_2 + \ddot{\lambda}\, a_2 + \dot\lambda a_1 \dot\lambda a_3 +  \dot\lambda a_4 \dot\lambda  a_2)/(\dot{\lambda}(a_2 - a_3)),$$ $$\mu(t) \equiv\dot{\lambda} \,\frac{(a_2 - a_3)}{\cu\,\omega_B^2},$$
remove momentum-dependent S-B couplings in $H_{CD}$, yielding the Hamiltonian 
\begin{align}
   H_{2} &= \frac{1}{2\,M(t)}P^2 + \frac{\tilde{\Lambda}(t)}{2} X^2 + \frac{1}{2}p_B^2  + \frac{K'(t)}{2}x_B^2 \nonumber \\
& - \cu\, \omega_B^2 \,x_B\, X  -\,\frac{1}{2}\,\big( \eta(t) + \mu(t)\,\tilde{\Lambda}(t) \big) (X\,P + P\,X). \nonumber
\end{align}
Here $$\bar{\Lambda}(t)\equiv \omega_S^2 - (\dot{\lambda} a_1)^2 - \partial_t (a_1) - (\dot{\lambda} \,a_2)^2 + \eta^2 + \partial_t \eta,$$ $$ K'(t) \equiv \omega_B^2 + \dot{\lambda} a_2 \, ( \dot{\lambda} a_2 -  2\dot{\lambda} a_3)- (\dot{\lambda} a_4)^2 - \partial_t (\dot{\lambda} a_4),$$ and  $$M^{-1}(t) \equiv 1+2\,\eta\,\mu + \bar{\Lambda}\,\mu^2 - \partial_t \mu .$$
Note that these transformations also shift the squared-frequency of the system and bath modes, generate a unit-less mass $M(t)$, and produce a term proportional to the dilation operator $\sim (XP+PX)$ of the system. 

The extra mass and dilation terms can be removed using the same transformations that appear in the construction of a FF protocol for a single dilated harmonic oscillator in vacuum~\cite{Deffner,Kolodrubetz}. The transformation $U_3(t)$ is a canonical re-scaling of $X$ and $P$, so that $(M,\tilde{\Lambda}) \to (1,\tilde{\Lambda}\, M^{-1})$. The transformation $U_4(t)$, shifts momentum to remove the term $\sim (XP+PX)$:
$$ U_{3} = \exp\bigg(  i \frac{\ln[M]}{4} \,\{X, P\} \bigg), \quad U_{4} = \exp\bigg( - i \,\xi\,\frac{X^2}{2} \bigg),$$
where $$ \xi(t) \equiv \eta + \mu\, \bar{\Lambda} + \frac{1}{2} \partial_t \ln[M] .$$
These unitary transformations yield the FF Hamiltonian $H_{FF}^{'} = H_4$:
\be \label{HFF}
H_{FF}^{'} = \frac{P^2}{2}  + \frac{\Lambda'(t)\,X^2}{2} + \frac{p_B^2}{2} + \frac{K'(t)\,x_B^2}{2} - C'(t) \,x_B\, X
 \ee
 where $$\Lambda'(t) \equiv \bar{\Lambda}\,M^{-1} - \xi^2 - \partial_t \xi,$$ $$C'(t) \equiv \cu\,\omega_B^2 \, \sqrt{M^{-1}}. $$ In what follows, we denote $K'(t) \equiv \omega_B^2 + z^2(t)$, where $$z^2(t) \equiv \dot{\lambda} a_2 \, ( \dot{\lambda} a_2 -  2\dot{\lambda} a_3)- (\dot{\lambda} a_4)^2 - \partial_t (\dot{\lambda} a_4). $$

\noindent
\textbf{(ii) Floquet-Engineered Fast Forward Drive.} The FF protocol in equation (\ref{HFF}) can be implemented by controlling only the system's frequency and a local coupling to the environment. The term $K'(t)$ cannot be manipulated directly, but can be effectively engineered by applying an additional Floquet modulation of the system-bath coupling. Then $K'(t)$ appears in the leading order of a high-frequency Magnus expansion of a Floquet Hamiltonian.  

Consider the Hamiltonian 
\begin{align}\nonumber
 H_{FF}^{FE}=\frac{1}{2}P^2& + \frac{1}{2} \,[\Lambda'(t) - z^2(t)] \,X^2 + \frac{p_B^2}{2} + \frac{\omega_B^2\,x_B^2}{2} \\ & - [ C'(t) - \sqrt{2}\,z(t)\, \Omega \cos(\Omega \,t)  ] \,x_B\,X 
 \label{eq:tilde_H}
\end{align}
 The Floquet frequency $\Omega$ is taken to be large enough to allow for a timescale separation between oscillatory part of the drive ($\cos(\Omega t)$) and all other time-dependent parameters ($\Lambda'(t), C'(t), z(t)$). These parameters then become effectively constant on the timescale of the Floquet driving period.

A simple way to find the Floquet Hamiltonian in this system is to go to the rotating frame with respect to the oscillating term~\cite{Bukov2}. To leading order in $1/\Omega$, we have 
\[
H_F\approx \overline{\mathrm e^{-i\sqrt{2} z \sin(\Omega t)\, x_B\, X} H' \mathrm e^{i\sqrt{2} z \sin(\Omega t)\, x_B\, X}},
\]
where the overline stands for period averaging and $H'$ is the Hamiltonian~\eqref{eq:tilde_H} without the oscillating term. For harmonic systems, the time averaging is easy to compute and only the kinetic energy terms generate new terms not present in $H'$:
\[
\overline{\mathrm e^{-i\sqrt{2} z \sin(\Omega t)\, x_B\, X}\, \frac{P^2}{2}\, \mathrm e^{i\sqrt{2} z \sin(\Omega t)\, x_B\, X}}=\frac{P^2}{2}+\frac{z^2 x_B^2}{2}
\]
and similarly for $p_B^2/2$. The effective Floquet Hamiltonian then reads
\begin{align}
H_F\approx \frac{P^2}{2} + \frac{\Lambda'(t)}{2}  \,X^2 + \frac{p_B^2}{2}
+ \frac{K'(t)}{2}\,x_B^2 - C'(t) \,x_B\,X 
\end{align}
where we have used $z^2(t)=K'(t)-\omega_B^2$. Therefore in the high frequency limit ($\Omega\to\infty$), $H_F$ becomes equivalent to $H_{FF}'$ in Eq.~\eqref{HFF}. 

A few comments are in order. First, the Floquet-Engineered FF Hamiltonian is only defined when $z^2(t) > 0$. This condition is generally satisfied for the protocols $\lambda(t)$ considered in this work. Second, the period averaging is sensitive to a gauge choice of the interval over which the period is measured~\cite{Bukov2}. This implies the dynamics of $H_F$ and $H_{FF}^{FE}$ are stroboscopically equilvalent, i.e. their evolution operators are identical only at integer multiples of the period. It follows that $H^{'}_{FF}$ and $H_{FF}^{FE}$ yield the same dynamics stroboscopically in the high-frequency limit.  

The equivalence of the dynamics of $H^{'}_{FF}$ and  $H_{FF}^{FE}$ at high-frequencies enables us to achieve a fast-forward protocol which implements $H_{CD}$ with the accessible experimental controls
\begin{align}
\tilde{\omega}^2_S(t) &= \Lambda'(t) - z^2(t) \\
\omega_B^2\,\tcu(t) &=  C'(t) - \sqrt{2}\,z(t)\, \Omega \cos(\Omega \,t), 
\end{align}
given any bare protocol $\lambda(t)$ satisfying proper boundary conditions. 

\begin{figure}[ht]
\centering
  \includegraphics[width=1.06\columnwidth]{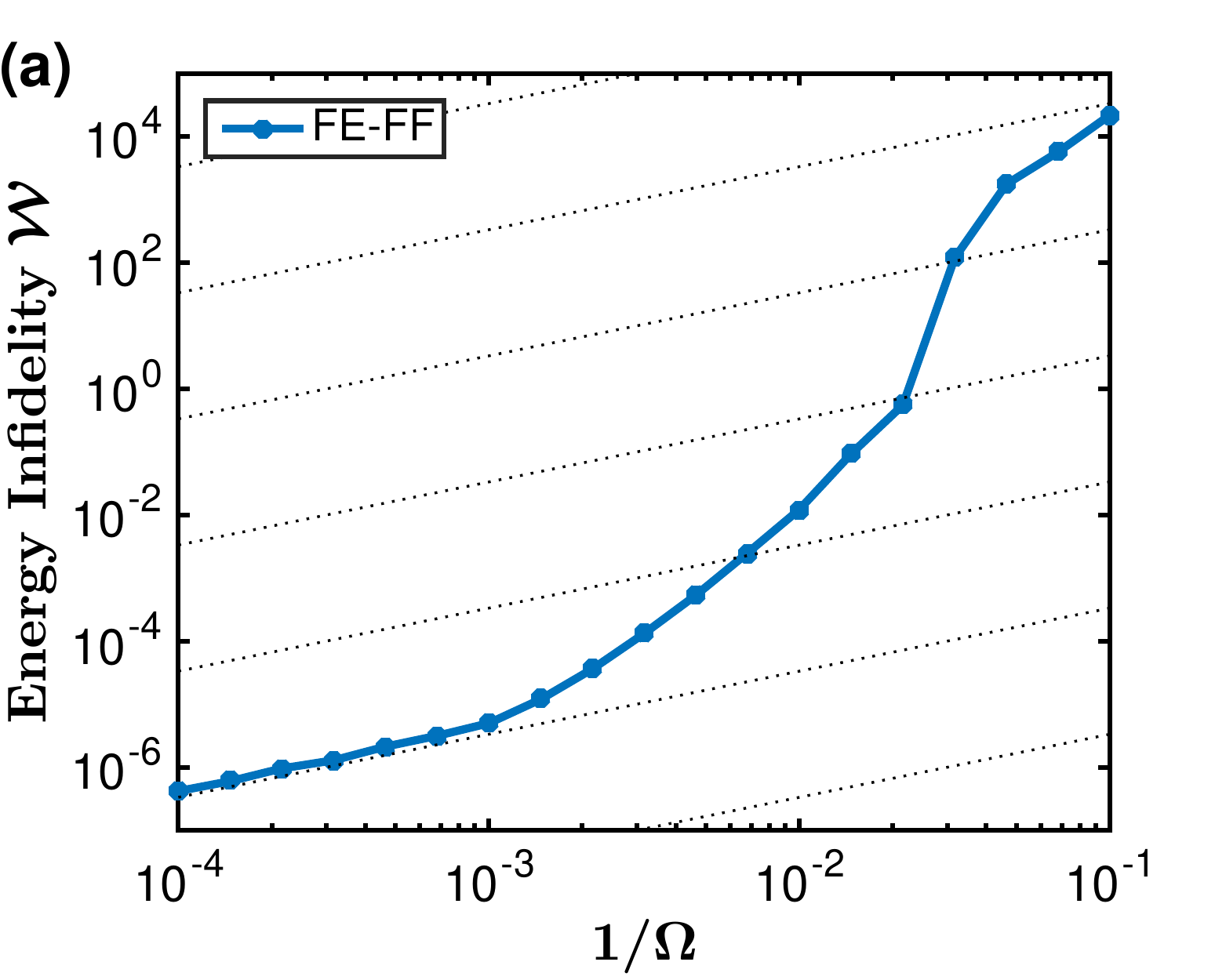}
  \includegraphics[width=1.06\columnwidth]{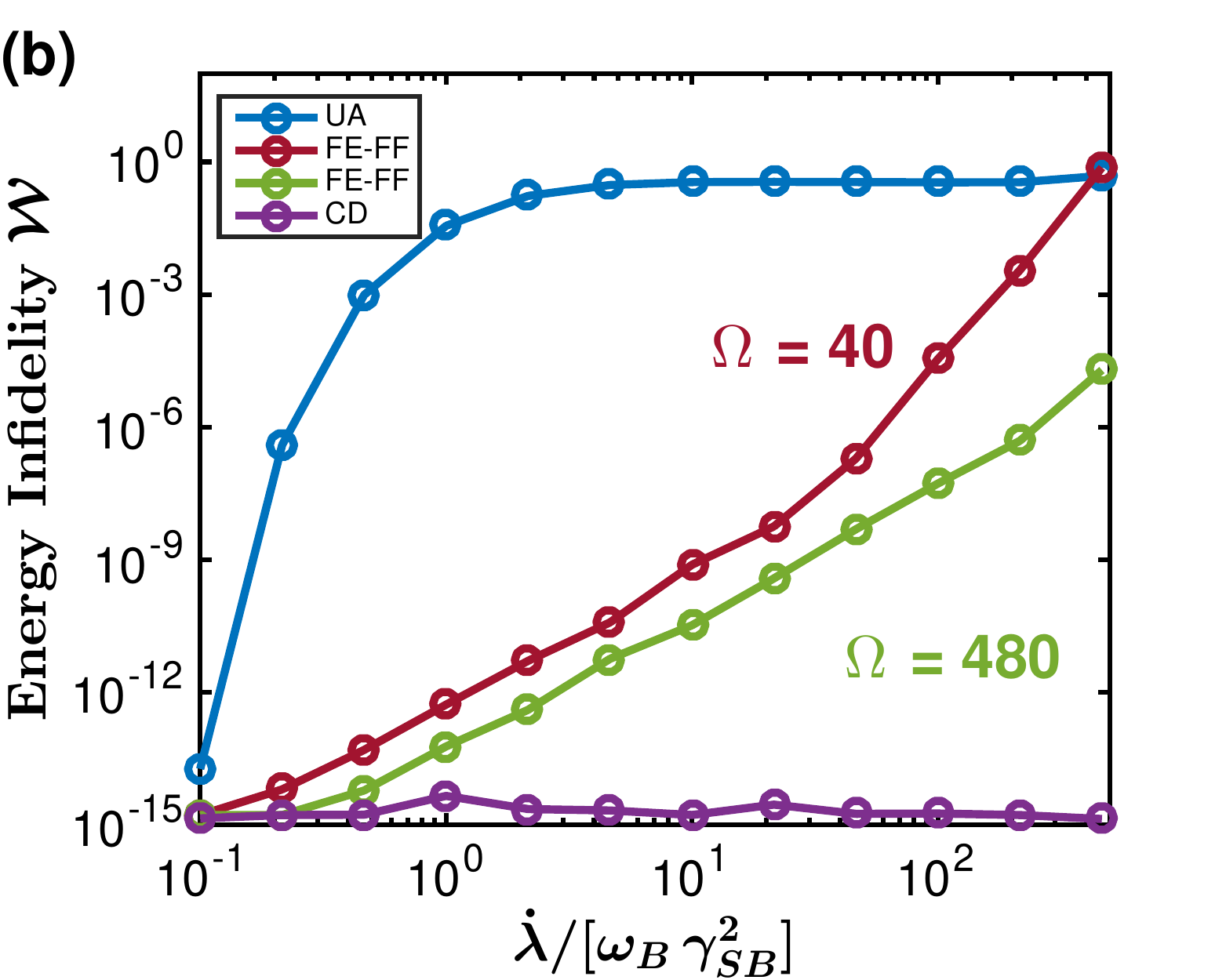}
    \caption{\textbf{Increasing the drive frequency $\Omega$ of a FE-FF protocol minimizes diabatic transitions.} (a) Simulation results for the energy infidelity $\mathcal{W}$ (c.f. Eq.~(9) from the main text) as a function of the inverse drive frequency $1/\Omega$, for a fast ramp with $\dot{\lambda} = 500\,\omega_B\,\cu^2$. The plot shows the convergence $\mathcal{W} \sim \Omega^{-1} \to 0$ as $\Omega\to\infty$, in agreement with the high-frequency equivalence of $H^{FE}_{FF}$ and $H_{CD}$. (b) Energy infidelity $\mathcal{W}$ as a function of normalized ramp speed $\dot{\lambda}/[\omega_B\,\cu^2]$. The plot shows how FE-FF protocols suppress diabatic transitions by several orders of magnitude compared to UA, when $\Omega$ is much larger than all other relevant frequency scales. Simulation Parameters: $\lambda_i=-0.67$, $\lambda_f=+0.67$, $\omega_B^2 = 3$, $\cu = 0.02$, and $\gu=0$.}
  \label{fig:FEConv}
\end{figure}

In Fig.~\ref{fig:FEConv}a, we demonstrate the performance of the Floquet-Engineered FF protocol by plotting the energy infidelity $\mathcal{W}$ (c.f. Eq.~(9) of the main text) as a function of the inverse frequency $\Omega^{-1}$. The dotted lines are chosen to have unit slope. The plot shows that $\mathcal{W}\sim \Omega^{-1}$ as $\Omega\to\infty$. Fig.~\ref{fig:FEConv}b shows how a high-frequency FE-FF protocol can decrease the energy infidelity by several orders of magnitude compared to UA, over a whole range of speeds $\dot{\lambda}$ spanning several decades.  \\

\subsection{Appendix E: Engine }\label{sec: AppE}

This section describes the application of FF driving to speed up thermalization processes in heat engines. A detailed description of the engine is given in the main text. \\

\noindent
\textbf{Slow ramp speeds:} As $\dot{\lambda} \to 0$, the UA and FF protocols coincide. The time evolution of S under UA and FF protocols is thus nearly identical at slow ramp speeds.

In the forward ramp $(\dot{\lambda}>0)$, S comes into contact with the cold bath C when its frequency is $\omega_S = \omega_C (1-\gu)$.
S thermalizes to the temperature of the cold bath $T_C$ at this point. 
It then undergoes an isothermal process at temperature $T_C$ as its frequency sweeps across the bandwidth of the cold bath, i.e. between $\omega_S=\omega_C (1-\gu)$ and $\omega_S=\omega_C (1+\gu)$.
Once $\omega_S > \omega_C (1+\gu)$, S is effectively isolated and contracts adiabatically until the point where $\lambda(t)$ is reversed.
In the backward ramp $(\dot{\lambda}<0)$, S expands adiabatically until its frequency coincides with the edge of the hot bath's bandwidth, i.e. until $\omega_S=\omega_H (1+\gu)$. 
At this point, S thermalizes to the temperature of the hot bath $T_H$.
It then undergoes an isothermal process at temperature $T_H$ as its frequency is decreased across the bandwidth of the hot bath, i.e. as $\omega_S$ is reduced from $\omega_H (1+\gu)$ to $\omega_H (1-\gu)$.
Once $\omega_S < \omega_H (1-\gu)$, S expands adiabatically until it returns to its initial configuration.
This cycle is schematically depicted by the solid curves in Fig. 5 of the main text. 

In the slow ramp speed limit, it is straightforward to calculate the heat absorbed (emitted) from baths H (C). 
When S thermalizes at the edge of the cold bath bandwidth at $\omega_S = \omega_C(1-\gu)$, its average occupation changes from $\langle n_i \rangle = T_H/[\omega_H(1-\gu)]$ to $\langle n_f \rangle = T_C/[\omega_C(1-\gu)]$. Thus, the heat emitted to the cold bath is: $$ Q_C^{th} = \omega_C \bigg( \frac{T_H}{\omega_H} -\frac{T_C}{\omega_C} \bigg) .$$ The heat ejected into the cold bath from the subsequent isothermal process is given by the integral of $T_C/\omega_S$ over the bandwidth of C. Therefore, the total heat ejected into C is:
\begin{align} \nonumber
 Q_C &= Q_C^{th} + \int_{\omega_{C}(1-\gu)}^{\omega_{C}(1+\gu)} \frac{T_C}{\omega_S} \,d\omega_S \\\label{QC0}
  &=\omega_C \bigg( \frac{T_H}{\omega_H} -\frac{T_C}{\omega_C} \bigg) + T_C \ln\bigg(\frac{1+\gu}{1-\gu}\bigg) 
 \end{align}
Similarly,
\begin{align} \nonumber
 Q_H &= Q_H^{th} + \int_{\omega_{H}(1-\gu)}^{\omega_{H}(1+\gu)} \frac{T_H}{\omega_S} \,d\omega_S \\ \label{QH0}
  &=\omega_H \bigg( \frac{T_H}{\omega_H} -\frac{T_C}{\omega_C} \bigg) + T_H \ln\bigg(\frac{1+\gu}{1-\gu}\bigg) .
 \end{align}
The efficiency and power obtained from expressions ($\ref{QC0}$) and ($\ref{QH0}$) are given in Eq.~(10) of the main text. Note that $P\to0$, since $\tau\to\infty$ in the slow limit.

The thermalization process at the edge of the H/C bath bandwidth makes the cycle irreversible.  Consequently, the efficiency is less than the Carnot bound $\eta_C= 1-T_C/T_H$. To attain the Carnot bound, we must impose the reversibility condition 
\be
\frac{T_C}{T_H} = \frac{\omega_C}{\omega_H} \quad\quad \text{or} \quad\quad r \equiv \frac{T_C}{T_H}\frac{\omega_H}{\omega_C} = 1
\ee
so that $Q_{C,H}^{th}=0$ at $r=1$. The efficiency is then: 
\be
\eta = 1-\frac{Q_C}{Q_H} = 1 - \frac{T_C}{T_H} \frac{\ln\big(\frac{1+\gu}{1-\gu}\big)}{\ln\big(\frac{1+\gu}{1-\gu}\big)}  = \eta_C.
\ee

\noindent
\textbf{Fast Driving.} The FF drive boosts the performance of the engine in fast ramps. Assume $\dot{\lambda}$ is larger than all intrinsic frequency scales, in particular, the thermalization rates $\omega_B \gu$ and the interaction rates $\omega_B \cu$ of both baths $H$ and $C$. We focus on the limit of $\gu \gg \cu$ below.  

Consider the energy change in either bath due to the resonant S-B exchange [$(x_B,p_B) \to -(X,P)$]:
\begin{align}\nonumber
Q \equiv& |\Delta\langle H_{bath} \rangle| = \bigg|\, \omega_B \bigg(\langle n_S \rangle-\langle n_B \rangle\bigg)+ \\ \label{Qfast} &\gu\,\omega_B^2\,\langle (X + x_B) \, (x_{J+1} + x_{J-1}) \rangle \bigg|
\end{align} 
where $x_{J\pm1}$ denote the coordinates of the bath oscillators coupled on either side of B. $\langle n_S \rangle$ and $\langle n_B \rangle$ denote the average occupation numbers of S and B, respectively, before the switch.

The final state of B after the FF switch is uncorrelated with its neighbors because the initial state of S is uncorrelated with the bath. Therefore, $\langle X \, (x_{J+1} + x_{J-1}) \rangle = 0$. 

To evaluate $\langle x_B \, \,x_{J\pm1} \rangle$ we first express the bath oscillators $x_j$ in terms of their normal mode coordinates $\tilde{x}_k$:
\begin{align}
    x_j = \sqrt{\frac{2}{N+1}}\sum_{k=1}^{N}\, \sin\bigg( \frac{\pi\,k\,j}{N+1} \bigg)\, \tilde{x}_k
\end{align}
where we have used open boundary conditions.

Since the bath is initialized in a classical thermal state at temperature $T$, equipartition implies that $$ \langle \tilde{x}_{k}  \, \,\tilde{x}_{k'} \rangle =  \delta_{k,k'} \langle \tilde{x}_k^2 \rangle = \delta_{k,k'}\,T/\omega_k^2,$$ where the normal mode frequencies are obtained by the diagonalizing $H_{bath}$:
\be\label{BnmFreq}
\omega_k^2 = \omega^2_B \, \bigg( 1 - \,2\,\gu \cos\bigg(\frac{\pi k}{N+1}\bigg)  \bigg).
\ee
Therefore,  
\begin{align}\label{Xxb}
\langle x_B \, \,x_{J\pm1} \rangle &= T\,\sum_{k=1}^{N} \sin\bigg(\frac{\pi\,k\,J}{N+1}\bigg) \,\sin\bigg(\frac{\pi\,k\,(J\pm1)}{N+1}\bigg) \frac{1}{\omega_k^2} .
\end{align}
Using Eq.~\eqref{BnmFreq}, we evaluate Eq.~\eqref{Xxb} to leading order in $\gu$:
\begin{align}\label{Qfast1}
\gu\,\omega_B^2\,\langle x_B \, \,x_{J\pm1} \rangle&\approx T \,\gu^2.
\end{align}

A similar derivation, writing operators in terms of normal mode coordinates and expanding to leading order in $\gu$, gives
\be\label{Qfast2}
\langle n_B \rangle = \frac{T}{\omega_B}\,(1+\gu^2).
\ee

During engine cycles, S alternates between swapping its occupation with a cold B oscillator and hot B oscillator. For example, after interacting with the hot bath, its occupation is given by 
\be\label{Qfast3}
\langle n_S \rangle = \frac{T_H}{\omega_H}\,(1+\gu^2).
\ee
This is the occupation of S before the subsequent switch with the cold bath. To obtain the heat transfer to the cold bath, we substitute Eqs.~\eqref{Qfast1}, \eqref{Qfast2} (with $T=T_C$), and \eqref{Qfast3} into \eqref{Qfast}: 
\begin{align}\label{Qgamma1}
Q_{C} = \omega_{C}\bigg(\frac{T_H}{\omega_H} - \frac{T_C}{\omega_C}\bigg)(1+\gu^2) + 2\, T_{C}\, \gu^2.
\end{align}

The heat absorbed from the hot bath can be derived by a similar argument:
\begin{align}\label{Qgamma2}
Q_{H} = \omega_{H}\bigg(\frac{T_H}{\omega_H} - \frac{T_C}{\omega_C}\bigg)(1+\gu^2) - 2\, T_{H}\, \gu^2.
\end{align}
Eqs.~\eqref{Qgamma1} and \eqref{Qgamma2} are summarized in Eq.~(13) of the main text. 

The efficiency is found to be
\be
\eta = 1 -\frac{\omega_C}{\omega_H}\bigg[\frac{(1-r)(1+\gu^2)+2\,r\,\gu^2}{(1-r)(1+\gu^2)-2\,\gu^2}\bigg]
\ee
where $r= [T_C \,\omega_H] / [T_H \,\omega_C]$. Observe that the efficiency is smaller than the $\gu = 0$ limit because less heat is drawn from H and more heat is dumped into C. 

The reversibility condition $r=1$ is no longer attainable since the engine fails at a sufficiently large $r=r_0<1$. The breakdown ratio $r_{0}$ is defined such that $Q_C = Q_H$, where the engine fails to extract useful work. Using equations (\ref{Qgamma1}) and (\ref{Qgamma2}) and expanding in $\gu$, we obtain 
\be
r_0 = 1 - 2\, \frac{(\omega_H+\omega_C)}{(\omega_H-\omega_C)}\,\gu^2.
\ee
We therefore operate the engine at $r<r_0$ to extract useful work as high ramp speeds. 

\begin{figure}[ht]
\centering
  \includegraphics[width=1.06\columnwidth]{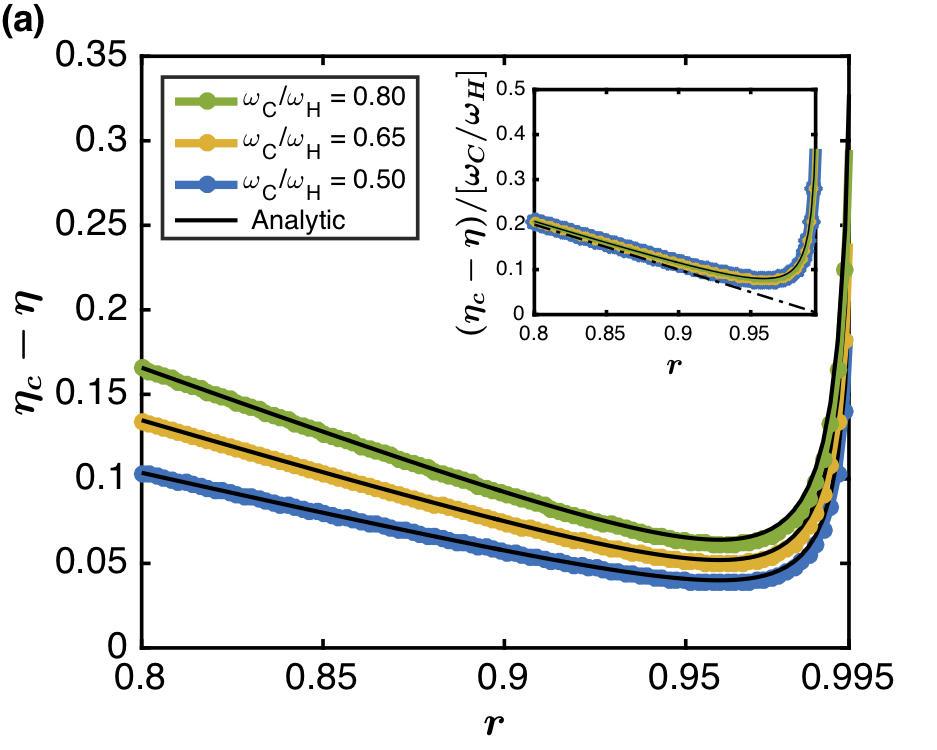}
  \includegraphics[width=1.06\columnwidth]{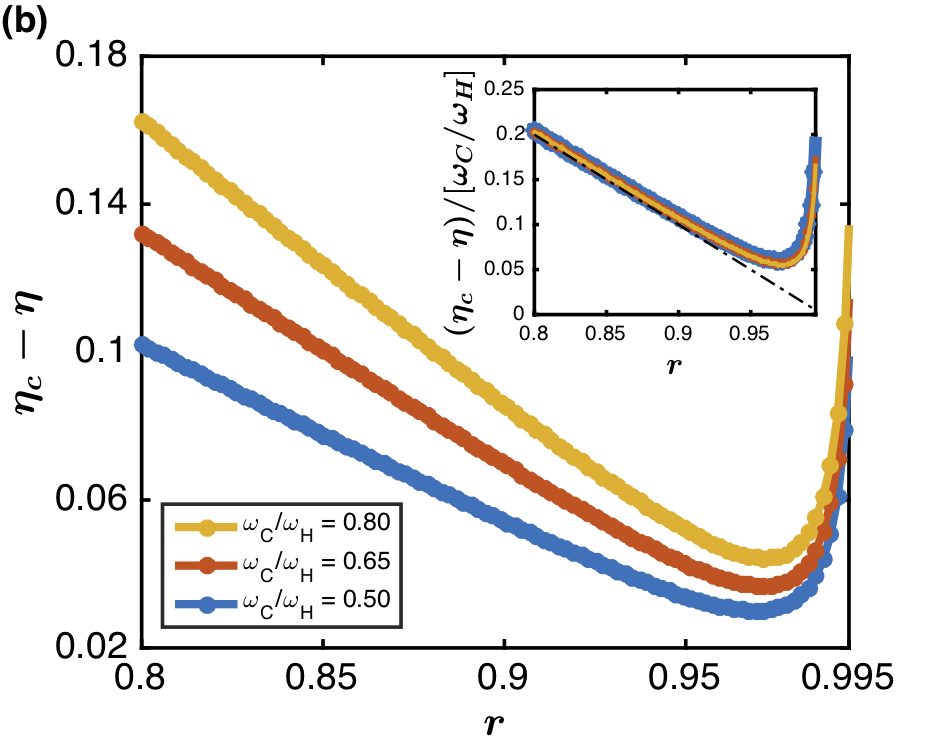}
    \caption{\textbf{Deviation of the efficiency from the Carnot bound as a function r  $\boldsymbol{=(T_C/T_H)/(\omega_C/\omega_H)}$.} (a) Simulation data for $\gu \gg \cu$, together with analytic curves obtained from equation (\ref{etaDiff}). Here $r_{min} \approx 0.96$ and $r_0 > 0.995$. The inset shows the curves collapse onto each other upon re-scaling by $\omega_C/\omega_H$. (b) Simulated curves obtained in the same manner as in (a), now with $\gu=\cu$. Inset: The curves exhibit an approximate collapse upon re-scaling by $\omega_C/\omega_H$. The dot-dash lines in both insets show the ideal case with $\gu=0$. Simulation Parameters: (a-b) $\omega_H^2+\omega_C^2= 5$, $T_H = 100$, and $\dot{\lambda} \approx 0.2$. $r$ is obtained by varying $T_C$ from 80 to 99.5. (a) $\gu = 0.02$, $\cu = 0.01 \gu$. (b) $\gu=\cu=0.02$.}
  \label{fig:EffCollapse}
\end{figure}

There exists an optimal ratio $r=r_{min} < r_0$ which minimizes the deviation of $\eta$ from $\eta_C$. We minimize  
\be \label{etaDiff}
\eta_C-\eta = \frac{\omega_C}{\omega_H} \bigg[\frac{(1-r)^2\,(1+\gu^2)+4\,r\,\gu^2}{(1-r)\,(1+\gu^2)-2\,\gu^2} \bigg]
\ee
with respect to $r$ to obtain
\be
r_{min} = \frac{1-2\gu-\gu^2}{1+\gu^2} = 1 - 2\,\gu +\mathcal{O}(\gu^2).
\ee
The behavior of the efficiency as a function of $r$ is shown in Fig.~\ref{fig:EffCollapse}a. Observe that far from the reversibility condition $r\ll1$ the high-speed efficiency is comparatively different from $\eta_C$. Near $r_{min}$, $\eta$ is closest to $\eta_C$, and in fact, $\eta/\eta_C$ can be quite close to 1 (see for example Fig.~7 of the main text). For $r_{min} < r < r_0$ we see a sharp deviation of $\eta$ from $\eta_C$ are we approach the breakdown ratio $r_0$. The plot shows curves for different values $\omega_C/\omega_H$ which collapse upon re-scaling by $\omega_C/\omega_H$; see inset. This is expected from equation $(\ref{etaDiff})$ and emphasizes that the difference between $\eta$ and $\eta_C$ can always be made smaller by tuning the ratio $\omega_C/\omega_H$. For reference, the inset also shows a black dashed line representing the limit $\gu=0$, where it is possible to attain the Carnot efficiency at $r=1$ with zero power output. Away from $r=1$, the finite $\gu>0$ curves exhibit qualitatively similar behavior to the $\gu=0$ case. Only near $r=1$ do we see significant deviations from the $\gu=0$ case, as the irreversible heat terms $\mathcal{O}(\gu^2)$ in equations (\ref{Qgamma1}) and (\ref{Qgamma2}) dominate the exchange. 

While we have focused on $\gu \gg \cu$ for simple analytic derivations, these results can be generalized to $\gu \gtrsim \cu$ by including $\cu$ corrections. The treatment is more involved since we must take into account the finite extent of the FF S-B exchange in the frequency domain (that is, the exchange no longer occurs at resonance, but over a frequency domain around resonance). Nevertheless, the behavior for $\gu \gtrsim \cu$ has been studied numerically in Fig~\ref{fig:EffCollapse}b and has been found to be of the same qualitative nature as $\gu \gg \cu$.

\subsection{Appendix F: Simulations}\label{sec: AppF}

\textbf{Simulations.} We simulate the dynamics of $(N+1)$ coupled oscillators in the Heisenberg picture.
Specifically, we numerically solve the Heisenberg equations of motion for the normal-mode creation/annihilation operators $(b_1(t),b_1^{\dagger}(t) , b_2(t), b_2^{\dagger}(t), ...)$ in the basis $(b_1(0),b_1^{\dagger}(0) , b_2(0), b_2^{\dagger}(0), ...)$ at $t=0$. Here $N$ denotes the number of bath oscillators. We take as input the parameters $\omega_B$, $\cu$, and $\gu$, as well as the ramp parameters described next.

The ramp protocol $\lambda(t)$ takes in an initial value $\lambda_i = \lambda(0) < 0$ at $t_i=0$, a final value $\lambda_f = \lambda(t_f) > 0$, a ramp up/down interval $\delta \lambda \ll \lambda_f - \lambda_ i$, and a maximum ramp speed $\dot{\lambda} = \dot{\lambda}_0$. The speed $\dot{\lambda}$ is increased from $0$ to $\dot\lambda_0$ for $\lambda$ in the interval [$\lambda_i$,$\lambda_i+\delta \lambda$] following a polynomial smoothstep of sixth order. 
The ramp is linear with $\dot{\lambda} = \dot{\lambda}_0$ from $\lambda = \lambda_i + \delta \lambda$ to $\lambda =  \lambda_f - \delta \lambda$. 
In particular, the ramp is linear at resonance.
The subsequent ramp-down of $\dot{\lambda}$ to zero also follows a polynomial smoothstep of sixth order over an interval [$\lambda_f - \delta \lambda$, $\lambda_f$]. 
The ramp up/down intervals are necessary to satisfy boundary conditions (see Appendix D). 
In the text, $\dot{\lambda}_0$ is the speed of the ramp.

The initial conditions used in simulations depend on the application. 
When $\gu = 0$, we initialize the S-B system in an eigenstate $|n_-(0),n_+(0) \rangle$ of the 2-oscillator Hamiltonian $H_0(\lambda_i)$ and compare the time evolved state to the adiabatically connected eigenstate $|n_-(t_f),n_+(t_f) \rangle$ of $H_0(\lambda_f)$.
When $\gu > 0$, S is connected to a 1d chain that models an optical phonon bath at temperature $T$. 
In this case, the bath normal-mode occupations are initialized in their corresponding high temperature Gibbs distributions with expectation values $\langle n_j \rangle = T/\omega_j$. 
Since S is far from resonance at $t=0$, it is essentially an independent normal-mode.
We therefore initialize it separately at a temperature different from the bath.  

To simulate the engine, we perform two ramps: a forward ramp $\dot{\lambda}>0$ as described above, and a backward ramp $\dot{\lambda}<0$ which runs in reverse. In each cycle, we must disconnect S from a cold/hot bath and connect it to the hot/cold bath. The connecting/disconnecting operations must be done slowly enough to avoid generating excess heat, or sufficiently far from resonance that this excess heat becomes negligible. This process is easily sped-up by using a different CD/FF protocol to turn on/off the coupling $\cu$ away from resonance. \\

\newpage
\afterpage{\blankpage}

\bibliographystyle{unsrtnat}
\bibliography{References}

\end{document}